\documentclass[journal]{IEEEtran}

\usepackage{color}
\usepackage{lineno}
\usepackage{cite}
\usepackage{verbatim}
\usepackage{pifont}
\usepackage{longtable}
\usepackage{tabularx}  
\usepackage{ltxtable}  
\usepackage{stfloats}
\usepackage{amsmath,amssymb,amsfonts}
\usepackage{algorithmic}
\usepackage{arydshln}
\usepackage{graphicx}
\usepackage{textcomp}
\usepackage{subfigure}
\usepackage{setspace}
\usepackage{svg}
\usepackage[ruled]{algorithm2e}
\usepackage{booktabs}
\usepackage{url}
\usepackage{diagbox}
\usepackage{multicol}
\usepackage{multirow}
\usepackage{tikz}% Make Orcid icon
\usepackage[colorlinks=true,linkcolor=blue,citecolor=blue,urlcolor=blue]{hyperref}
\definecolor{lime}{HTML}{A6CE39}
\DeclareRobustCommand{\orcidicon}{%
    \begin{tikzpicture}
    \draw[lime, fill=lime] (0,0) 
    circle [radius=0.16] 
    node[white] {{\fontfamily{qag}\selectfont \tiny ID}};    \draw[white, fill=white] (-0.0625,0.095) 
    circle [radius=0.007];    \end{tikzpicture}
    \hspace{-2mm}}
\foreach \x in {A, ..., Z}{%
    \expandafter\xdef\csname orcid\x\endcsname{\noexpand\href{https://orcid.org/\csname orcidauthor\x\endcsname}{\noexpand\orcidicon}}
    }

\usepackage{listings}
\usepackage{xcolor}

\definecolor{codegreen}{rgb}{0,0.6,0}
\definecolor{codegray}{rgb}{0.5,0.5,0.5}
\definecolor{codepurple}{rgb}{0.58,0,0.82}
\definecolor{backcolour}{rgb}{0.96,0.96,0.96} % Slightly off-white background

\lstdefinestyle{matlabstyle}{
    backgroundcolor=\color{backcolour},   
    commentstyle=\color{codegreen},
    keywordstyle=\color{blue},
    numberstyle=\tiny\color{codegray},
    stringstyle=\color{codepurple},
    basicstyle=\ttfamily\scriptsize, % Use small typewriter font
    breakatwhitespace=false,         
    breaklines=true,                 % Auto line breaking
    captionpos=b,                    
    keepspaces=true,                 
    numbers=left,                    % Line numbers on the left
    numbersep=5pt,                  
    showspaces=false,                
    showstringspaces=false,
    showtabs=false,                  
    tabsize=4,
    language=Matlab,
    frame=lines,                     % Top and bottom lines
    rulecolor=\color{black}
}

\SetCommentSty{mycommfont}

\SetKwInput{KwInput}{Input}                % Set the Input
\SetKwInput{KwOutput}{Output}              % set the Output

\begin{document}

% Line numbers generation.
% \linenumbers
% \pagewiselinenumbers
% \switchlinenumbers

% Generation for 3.0x line width.
% \begin{spacing}{3.0}·

% Paper title and information.
\title{Represent Micro-Doppler Signature in Orders\\
\thanks{Manuscript received XXXXXXX XX, 2026; revised XXXXXXX XX, 2026; accepted XXXXXXX XX, 2026. Date of publication XXXXXXX XX, 2026; date of current version XXXXXXX XX, 2026.\par
My Bio: My name is Weicheng Gao. I'm a Ph.D. student from Beijing Institute of Technology. I’m majored and interested in mathematical and modeling theory research of signal processing, radar signal processing techniques, and AI for radar, apprenticed under professor Xiaopeng Yang. I’m currently dedicated in the field of Through-the-Wall Radar Human Activity Recognition. Looking forward to learning and collaborating with more like-minded teachers and mates. (e-mail: JoeyBG@126.com).\par
Digital Object Identifier 10.48550/arXiv.2602.XXXXX.\par}}

% Author definitions.
\author{Weicheng~Gao\orcidA{},~\IEEEmembership{Graduate~Student~Member,~IEEE}   
        % <-this % stops a space
        \vspace{-0.4cm}
        }
        
% The paper headers.
\markboth{arXiv Preprint, February, 2026}%
{Shell \MakeLowercase{\textit{et al.}}: Bare Demo of IEEEtran.cls for IEEE Journals}

% Make the title area.
\maketitle

%%%%%%%%%%%%%%%%%%%%%%%%%%%%%%%%%%%%%%%%%%%%%%%%%%%%%%%%%%%%%%%%%%%%%%%%%%%%%%%%%%
% Abstract
%%%%%%%%%%%%%%%%%%%%%%%%%%%%%%%%%%%%%%%%%%%%%%%%%%%%%%%%%%%%%%%%%%%%%%%%%%%%%%%%%%
\begin{abstract}
Non-line-of-sight sensing of human activities in complex environments is enabled by multiple-input multiple-output through-the-wall radar (TWR). However, the distinctiveness of micro-Doppler signature between similar indoor human activities such as gun carrying and normal walking is minimal, while the large scale of input images required for effective identification utilizing time-frequency spectrograms creates challenges for model training and inference efficiency. To address this issue, the Chebyshev-time map is proposed in this paper, which is a method characterizing micro-Doppler signature using polynomial orders. The parametric kinematic models for human motion and the TWR echo model are first established. Then, a time-frequency feature representation method based on orthogonal Chebyshev polynomial decomposition is proposed. The kinematic envelopes of the torso and limbs are extracted, and the time-frequency spectrum slices are mapped into a robust Chebyshev-time coefficient space, preserving the multi-order morphological detail information of time-frequency spectrum. Numerical simulations and experiments are conducted to verify the effectiveness of the proposed method, which demonstrates the capability to characterize armed and unarmed indoor human activities while effectively compressing the scale of the time-frequency spectrum to achieve a balance between recognition accuracy and input data dimensions. The open-source code of this paper can be found in: \href{https://github.com/JoeyBGOfficial/Represent-Micro-Doppler-Signature-in-Orders}{GitHub/JoeyBGOfficial/Represent-Micro-Doppler-in-Orders}.\par
\end{abstract}


\IEEEpeerreviewmaketitle

%%%%%%%%%%%%%%%%%%%%%%%%%%%%%%%%%%%%%%%%%%%%%%%%%%%%%%%%%%%%%%%%%%%%%%%%%%%%%%%%%%
% Introduction
%%%%%%%%%%%%%%%%%%%%%%%%%%%%%%%%%%%%%%%%%%%%%%%%%%%%%%%%%%%%%%%%%%%%%%%%%%%%%%%%%%
\section{Introduction}
\IEEEPARstart{R}{esearch} on through-the-wall radar (TWR) human activity recognition (HAR) holds significant theoretical value and practical importance \cite{Amin}. This technology enables non-contact, high-penetration covert detection, providing critical information for search and rescue operations and situational awareness in counter-terrorism and disaster relief scenarios \cite{CuiGL, GuoSS, JiaY, JinT}. Furthermore, it promotes frontier breakthroughs in radar signal processing and pattern recognition, offering innovative means for ensuring life safety in complex environments \cite{DingYP, AnQ, YangDG, YeSB}.\par
\begin{figure}
    \centering
    \includegraphics[width=0.48\textwidth]{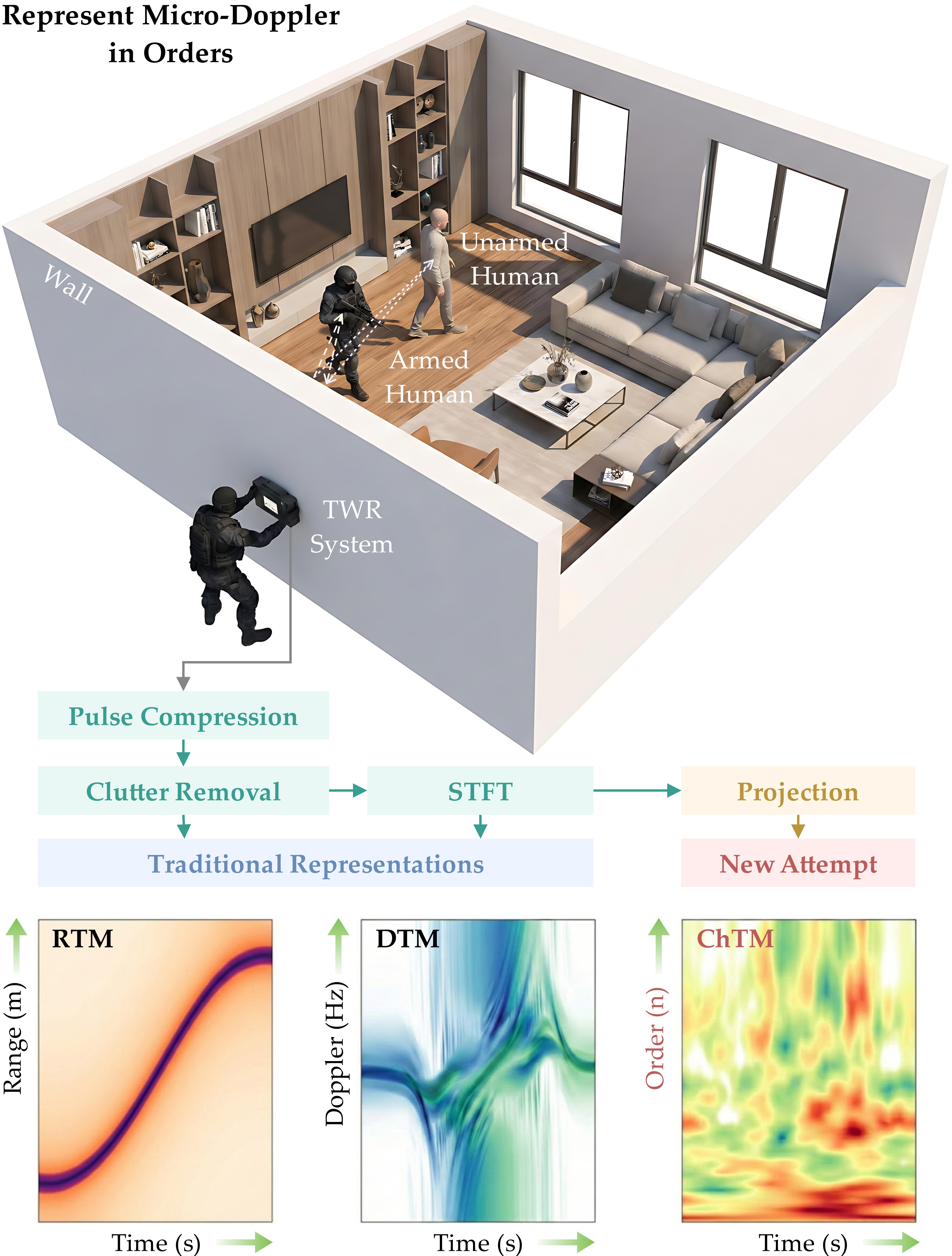}
    \caption{Overall concept of representing micro-Doppler in orders.}
    \label{Overall}
    \vspace{-0.4cm}
\end{figure}\par
Over the past five years, the work on HAR utilizing radar has been gradually matured \cite{Review1, Review2, Review3}. Except that the center frequency, bandwidth, and clutter suppression methods of the utilized radar systems are different, the micro-Doppler signature extraction and recognition of TWR HAR are considered almost identical to those of traditional radar HAR \cite{KimY}. Among them, Yu et al. proposed a noninvasive millimeter-wave system that transformed raw radar returns into spatial-temporal point clouds and employed denoising, enhanced voxelization, data augmentation, and a dual-view convolutional neural network (CNN), which significantly improved activity classification and fall detection performance compared with conventional machine learning methods \cite{Yu2022}. Li et al. proposed a semisupervised joint domain and semantic transfer learning framework that exploited sparsely labeled micro-Doppler spectrograms to alleviate annotation demands, achieving high recognition accuracy even when only a small fraction of instances in the training dataset were labeled \cite{Li2021}. Huan et al. proposed an attention-based hybrid network that decoupled Doppler and temporal features by feeding sliding-window Doppler sequences into a one-dimensional CNN followed by an attention-enhanced long short-term memory (LSTM) module, which enhanced clutter suppression and yielded high recognition accuracy with a lightweight architecture suitable for embedded deployment \cite{Huan2023}. Tan et al. proposed a micro-Doppler HAR system that combined a two-stream one-dimensional CNN with a bidirectional gated recurrent unit and a cross-channel operation between magnitude and phase branches, which effectively captured spatial-temporal features from radar time–frequency responses and achieved superior accuracy on the Rad-HAR dataset\cite{Tan2024}. Ma et al. proposed a multi-directional and multi-scenario radar-based recognition system that constructed a small-sample dataset and designed HogEffNet, a lightweight feature enhancement and fusion network integrating EfficientNetB1, histogram of oriented gradients, transfer learning, and sparse autoencoder-based fusion, which improved robustness to viewpoint variation and reached state-of-the-art performance \cite{Ma2025}. Faisal et al. proposed a cross-terms free Wigner–Ville distribution method based on time segmentation and frequency windowing to obtain high-resolution time-frequency representations of frequency-modulated continuous-wave (FMCW) radar slow-time signature, and an ensemble classifier with optimized parameters was employed to recognize activities with very high accuracy\cite{Faisal2024}. Tang et al. proposed a mixed CNN that jointly exploited three radar time-frequency spectrograms obtained via short-time Fourier transform (STFT), reduced interference distribution with Hanning kernel, and smoothed pseudo Wigner-Ville distribution, where independent two-dimensional convolutional branches and a three-dimensional convolutional branch were fused to improve average recognition accuracy over existing methods\cite{Tang2021}. Huan et al. further proposed a lightweight hybrid vision transformer that combined efficient convolutional feature pyramids with Radar-ViT blocks using fold and unfold operations and RES-SE modules, which enhanced multi-scale micro-Doppler representations while maintaining low computational complexity for embedded HAR \cite{Huan2023ViT}. Zhu et al. proposed a continuous HAR framework based on distributed radar sensor networks, where two-dimensional CNN extracted spatial information from spectrograms and bidirectional gated recurrent units modeled temporal dependencies, and several neural-network-based fusion strategies for multiple radar nodes were investigated, resulting in high accuracy for unconstrained continuous activities \cite{Zhu2022}. Erol et al. proposed a multilinear subspace learning scheme that directly processed radar data cubes in slow-time, fast-time, and Doppler dimensions, optimized mode-wise subspaces to minimize reconstruction error, and combined multilinear principal component analysis with linear discriminant analysis and shallow neural networks in a boosting architecture, which achieved superior classification accuracy across subjects, positions, and aspect angles compared with spectrogram-based techniques \cite{Erol2019}. Besides, a series of works were proposed by our team in the field of TWR HAR \cite{Gao1, Gao2, Gao3, Gao4, Gao5, Gao6}. Accurate, robust, and efficient TWR HAR is achieved through previous work series with both simulated and measured experimental verifications. A common theoretical basis was shared by these works: Micro-Doppler visualization was achieved, features were extracted, and recognition was realized through the utilization of radar range profiles or time-frequency spectrograms \cite{Review1, Review2, Review3}. Although favorable recognition accuracy was achieved, the image representation of micro-Doppler signature was not pioneered.\par
For indoor HAR scenarios, significant differences in transient features and occupied Doppler bandwidths are exhibited by the micro-Doppler signature when different activities of the same individual are distinguished. However, minimal differences in micro-Doppler signature are observed when the same activity performed by different individuals is distinguished, or when whether a person is armed is determined. To achieve favorable recognition accuracy, the entire range-time map (RTM) or Doppler-time map (DTM) is input into the classifier, which is considered the mainstream concept at present. Micro-Doppler augmentation is achieved using multi-input multi-output (MIMO) TWR in our previous work, yet the scale of input data for processing algorithms is further increased by this approach \cite{Gao7}. Therefore, the development of a micro-Doppler image representation that is physically interpretable, preserves motion detail features, and compresses data scales is deemed necessary.\par
To address the issue mentioned above, as shown in Fig. \ref{Overall}, the Chebyshev-time map (ChTM) is proposed in this paper, which is a method characterizing micro-Doppler signature using polynomial orders. In this paper, a time-frequency feature representation method based on orthogonal Chebyshev polynomial decomposition is proposed. The kinematic envelopes of the torso and limbs are extracted, and the DTM slices are mapped into a robust Chebyshev-time coefficient space, preserving the multi-order morphological detail information of DTM. Detailed definitions and theoretical analyses are also included in this paper, and the physical background with the significance of the proposed ChTM micro-Doppler representation is proved. In addition, numerical simulations and experiments are conducted to verify the effectiveness of the proposed method, which demonstrates that the proposed ChTM compresses the scale of DTM to achieve a balance between recognition accuracy and input data dimensions.\par
The rest of the paper is organized as follows. The human kinematic model and corresponding TWR echo model are established in section II. The method of generating ChTM is proposed in section III with the proof of its physical background. Numerical simulations and experiments are conducted and analyzed in section IV. The conclusion of this paper is given in section V.\par

\section{Kinematic and TWR Echo Model}
In this section, the human armed and unarmed walking model, and the corresponding TWR echo model are first presented. Then, the method of achieving radar signal processing with RTM and DTM generation is discussed.\par

\subsection{Human Kinematic and TWR Echo Model}
\begin{figure}
    \centering
    \includegraphics[width=0.48\textwidth]{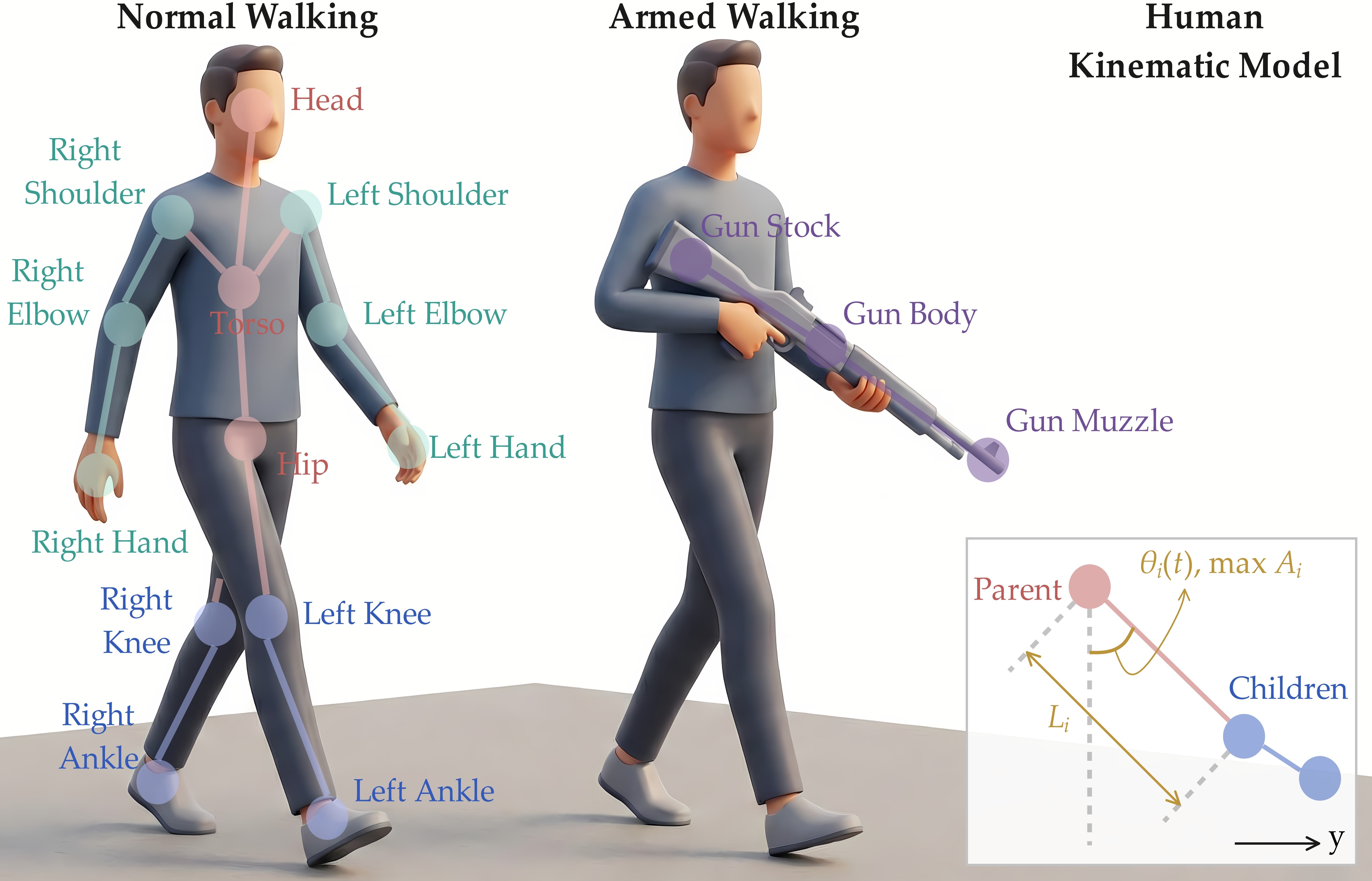}
    \caption{Human kinematic model for normal walking and armed walking.}
    \label{Human_Kinematic_Model}
    \vspace{-0.2cm}
\end{figure}

As shown in Fig. \ref{Human_Kinematic_Model}, based on Boulic-Thalmann human motion model \cite{Boulic}, the motion equations of each human limb joints can be established in the three-dimensional space.\par
Assume that the radar is located around $\mathbf{P}_{\mathrm{RadarCenter}}=(0,0,z_\mathrm{radar})^\top$. For MIMO radar, the coordinate denotes the center of the array. The human body moves in the $xOy$ plane. Let $t$ be the slow time.The human torso serves as the root joint, and its motion is determined by the initial position $\mathbf{P}_{\mathrm{init}} = [x_0, y_0, z_0]^\top$, the motion velocity $v_{\mathrm{torso}}$, and the motion direction angle $\phi_{\mathrm{walk}}$. The three-dimensional coordinates $\mathbf{P}_{\mathrm{Torso}}(t)$ of the torso center are:
\begin{equation}
\mathbf{P}_{\mathrm{torso}}(t) = \begin{bmatrix}
x_{\mathrm{Torso}}(t) \\
y_{\mathrm{Torso}}(t) \\
z_{\mathrm{Torso}}(t)
\end{bmatrix} = \begin{bmatrix}
x_0 + v_{\mathrm{torso}} \cos(\phi_{\mathrm{walk}}) \cdot t \\
y_0 + v_{\mathrm{torso}} \sin(\phi_{\mathrm{walk}}) \cdot t \\
h_{\mathrm{torso}}
\end{bmatrix},
\end{equation}
where $h_{\mathrm{torso}}$ is the height of the torso center.\par
The positions of other joints in the human body are recursively calculated through the position of the parent joint, the limb length vector $\mathbf{L}$, and the rotation matrix $\mathbf{R}(\theta(t))$. Let the connection vector of the $i$-th joint relative to its parent joint $p$ be $\mathbf{L}_i$, the joint rotation angle be $\theta_i(t)$, and the rotation axis be the $y$-axis. The instantaneous position of the $i$-th joint is:
\begin{equation}
\mathbf{P}_i(t) = \mathbf{P}_p(t) + \mathbf{R}_y(\theta_i(t)) \cdot \mathbf{L}_i,
\end{equation}
where the rotation matrix $\mathbf{R}_y(\theta)$ around the $y$-axis is:
\begin{equation}
\mathbf{R}_y(\theta) = \begin{bmatrix}
\cos\theta & 0 & \sin\theta \\
0 & 1 & 0 \\
-\sin\theta & 0 & \cos\theta
\end{bmatrix}.
\end{equation}\par
Define $f_\mathrm{gait}$ as the gait frequency. Below, two patterns including normal walking and armed walking are discussed in detail.\par
\textbf{(1) Normal Walking:}\par
In this pattern, the lower and upper limbs of the human body swing in opposite phases. Define the swing amplitudes of the thigh, calf, and arm as $A_{\mathrm{thigh}}, A_{\mathrm{calf}}, A_{\mathrm{arm}}$ respectively.\par
The derived joints of the torso include the head, right shoulder, and left shoulder. These three joints have fixed displacements relative to the torso and no rotations, where $\theta_{\mathrm{head}}(t),\theta_{\mathrm{rs}}(t),\theta_{\mathrm{ls}}(t)\equiv 0$, thus:
\begin{equation}
\begin{aligned}
\mathbf{P}_{\mathrm{Head}}(t) &= \mathbf{P}_{\mathrm{Torso}}(t) + [0, 0, \Delta z_{\mathrm{head}}]^\top\\
\mathbf{P}_{\mathrm{RS}}(t) &= \mathbf{P}_{\mathrm{Torso}}(t) + [0, -\Delta y_{\mathrm{sh}}, \Delta z_{\mathrm{sh}}]^\top\\
\mathbf{P}_{\mathrm{LS}}(t) &= \mathbf{P}_{\mathrm{Torso}}(t) + [0, \Delta y_{\mathrm{sh}}, \Delta z_{\mathrm{sh}}]^\top\\
\end{aligned},
\end{equation}
where $\Delta z_{\mathrm{head}},\Delta z_{\mathrm{sh}}$ are the offsets in $z$ axis of the head and the shoulders to the torso, respectively. $\Delta y_{\mathrm{sh}}$ is the offset in $y$ axis of the left shoulder to the torso.\par
The derived joints of the hip include the right knee, left knee, right ankle, and left ankle. The hip is fixed relative to the torso, where $\theta_{\mathrm{hip}}\equiv 0$, thus:
\begin{equation}
\mathbf{P}_{\mathrm{Hip}}(t) = \mathbf{P}_{\mathrm{Torso}}(t) + [0, 0, \Delta z_{\mathrm{hip}}]^\top,
\end{equation}
where $\Delta z_{\mathrm{hip}}$ is the offset in $z$ axis of the hip to the torso.\par
The knee joints are connected to the hip, and the ankle joints are connected to the knee, both performing periodic swings. The parent joint of the right knee is the hip. The swing angle of the right knee is $\theta_{\mathrm{rk}}(t) = A_{\mathrm{thigh}}\sin(2\pi f_{\mathrm{gait}} t)$, and:
\begin{equation}
\mathbf{P}_{\mathrm{RK}}(t) = \mathbf{P}_{\mathrm{Hip}}(t) + \mathbf{R}_y(\theta_{\mathrm{rk}}(t)) \cdot [0, 0, -L_{\mathrm{thigh}}]^\top,
\end{equation}
where $L_\mathrm{thigh}$ is the length of the thigh.\par
The parent joint of the right ankle is the right knee. The swing angle of the right ankle is $\theta_{\mathrm{ra}}(t) = A_{\mathrm{calf}}\sin(2\pi f_{\mathrm{gait}} t + \frac{\pi}{4})$, and:
\begin{equation}
\mathbf{P}_{\mathrm{RA}}(t) = \mathbf{P}_{\mathrm{RK}}(t) + \mathbf{R}_y(\theta_{\mathrm{ra}}(t)) \cdot [0, 0, -L_{\mathrm{calf}}]^\top,
\end{equation}
where $L_\mathrm{calf}$ is the length of the calf.\par
The parent joint of the left knee is also the hip. The swing angle of the left knee is the inversion of the right knee $\theta_{\mathrm{lk}}(t) = -A_{\mathrm{thigh}}\sin(2\pi f_{\mathrm{gait}} t)$, and:
\begin{equation}
\mathbf{P}_{\mathrm{LK}}(t) = \mathbf{P}_{\mathrm{Hip}}(t) + \mathbf{R}_y(\theta_{\mathrm{lk}}(t)) \cdot [0, 0, -L_{\mathrm{thigh}}]^\top.
\end{equation}\par
The parent joint of the left ankle is the left knee. The swing angle of the left ankle is the inversion of the right ankle $\theta_{\mathrm{la}}(t) = -A_{\mathrm{calf}}\sin(2\pi f_{\mathrm{gait}} t + \frac{\pi}{4})$, and:
\begin{equation}
\mathbf{P}_{\mathrm{LA}}(t) = \mathbf{P}_{\mathrm{LK}}(t) + \mathbf{R}_y(\theta_{\mathrm{la}}(t)) \cdot [0, 0, -L_{\mathrm{calf}}]^\top.
\end{equation}\par
The dynamic joints of the upper limbs connected to the shoulders include the right elbow, right hand, left elbow, and left hand. The arm swing is in the same direction as the contralateral leg. The parent joint of the right elbow is the right shoulder. The swing angle of the right elbow is $\theta_{\mathrm{re}}(t) = A_{\mathrm{arm}}\sin(2\pi f_{\mathrm{gait}} t + \pi)$, and:
\begin{equation}
\mathbf{P}_{\mathrm{RE}}(t) = \mathbf{P}_{\mathrm{RS}}(t) + \mathbf{R}_y(\theta_{\mathrm{re}}(t)) \cdot [0, 0, -L_{\mathrm{arm}}/2]^\top,
\label{Normal_RE}
\end{equation}
where $L_\mathrm{arm}$ is the length of the arm.\par
The parent joint of the right hand is the right elbow. There's no rotation between the hand and the elbow, thus $\theta_{\mathrm{rh}}(t)\equiv 0$, and:
\begin{equation}
\mathbf{P}_{\mathrm{RH}}(t) = \mathbf{P}_{\mathrm{RE}}(t) + [0, 0, -L_{\mathrm{arm}}/2]^\top.
\label{Normal_RH}
\end{equation}\par
The parent joint of the left elbow is the left shoulder. The swing angle of the left elbow is the inversion of the right elbow $\theta_{\mathrm{le}}(t) = -A_{\mathrm{arm}}\sin(2\pi f_{\mathrm{gait}} t + \pi)$, and:
\begin{equation}
\mathbf{P}_{\mathrm{LE}}(t) = \mathbf{P}_{\mathrm{LS}}(t) + \mathbf{R}_y(\theta_{\mathrm{le}}(t)) \cdot [0, 0, -L_{\mathrm{arm}}/2]^\top.
\end{equation}\par
The parent joint of the left hand is the left elbow. As the same, $\theta_{\mathrm{lh}}(t)\equiv 0$, and:
\begin{equation}
\mathbf{P}_{\mathrm{LH}}(t) = \mathbf{P}_{\mathrm{LE}}(t) + [0, 0, -L_{\mathrm{arm}}/2]^\top.
\end{equation}\par
\textbf{(2) Armed Walking:}\par
In this pattern, the motion equations for the lower limb joints are exactly the same as those during normal walking. The difference is that the upper limbs are locked in a gun-holding posture, and three gun joints are added.\par
The locked lower limb joints include the right elbow, right hand, left elbow, and left hand. The arm no longer swings over time but is fixed at a specific angle.
Let the fixed angle of the upper arm be $\theta_{\mathrm{upper}} = -\pi/6$, and the fixed angle of the forearm be $\theta_{\mathrm{fore}} = -4\pi/9$, thus:
\begin{equation}
\begin{aligned}
\mathbf{P}_{\mathrm{RE}}(t) &= \mathbf{P}_{\mathrm{RS}}(t) + \mathbf{R}_y(\theta_{\mathrm{upper}}) \cdot [0, 0, -L_{\mathrm{arm}}/2]^\top\\
\mathbf{P}_{\mathrm{RH}}(t) &= \mathbf{P}_{\mathrm{RE}}(t) + \mathbf{R}_y(\theta_{\mathrm{fore}}) \cdot [0, 0, -L_{\mathrm{arm}}/2]^\top\\
\mathbf{P}_{\mathrm{LE}}(t) &= \mathbf{P}_{\mathrm{LS}}(t) + \mathbf{R}_y(\theta_{\mathrm{upper}}) \cdot [0, 0, -L_{\mathrm{arm}}/2]^\top\\
\mathbf{P}_{\mathrm{LH}}(t) &= \mathbf{P}_{\mathrm{LE}}(t) + \mathbf{R}_y(\theta_{\mathrm{fore}}) \cdot [0, 0, -L_{\mathrm{arm}}/2]^\top
\end{aligned}.
\end{equation}\par
The gun is rigidly connected to the right hand and moves translationally with the right hand. Let the local offset vectors of the gun stock, gun body, and muzzle relative to the right hand be $\mathbf{d}_{\mathrm{stock}}, \mathbf{d}_{\mathrm{body}}, \mathbf{d}_{\mathrm{muzzle}}$ respectively, then:
\begin{equation}
\begin{aligned}
\mathbf{P}_{\mathrm{Stock}}(t) &= \mathbf{P}_{\mathrm{RH}}(t) + [0.1, 0, 0]^\top\\
\mathbf{P}_{\mathrm{Body}}(t) &= \mathbf{P}_{\mathrm{Stock}}(t) + [0.2, 0, 0]^\top\\
\mathbf{P}_{\mathrm{Muzzle}}(t) &= \mathbf{P}_{\mathrm{Body}}(t) + [0.3, 0, 0]^\top
\end{aligned}.
\end{equation}\par
The time-varying information of the three-dimensional coordinates of all human joints will be used for the calculation of the time delay of the scattering center in the subsequent TWR echo model \cite{RadHARSimulatorV1, RadHARSimulatorV2}.\par
\par

\subsection{TWR Echo Model}
Assume that the radar emits linear FMCW wave \cite{TWR_Echo}. Let the center frequency be $f_c$, the bandwidth be $B$, the pulse width be $T_p$, the frequency modulation slope be $K = B/T_p$, and the transmitting power be $P_\mathrm{tx}$. The transmitted signal $s_\mathrm{tx}(t_f)$ in the fast-time domain $t_f, 0 \le t_f \le T_p$ is:
\begin{equation}
s_\mathrm{tx}(t_f) = \sqrt{P_\mathrm{tx}} \cdot \exp\left( j 2\pi \left( f_c t_f + \frac{1}{2} K t_f^2 \right) \right).
\end{equation}\par
Consider that the wall is a single-layer homogeneous medium with a relative permittivity of $\epsilon_r$, a loss tangent of $\tan\delta$, and a thickness of $d_{\mathrm{wall}}$. Suppose that the $k$-th transmitting antenna is located at $\mathbf{T}_k$, the $l$-th receiving antenna is located at $\mathbf{R}_l$, and the position of the $i$-th scattering point is $\mathbf{P}_i(t)$. At the slow time $t$, the total path of the signal from the Tx to the target and then to the Rx includes the air segment and the in-wall segment. By adopting a simplified approximate optical path model, the distance from the Tx $\mathbf{T}_k$ to the target is $R_{\mathrm{tx},i} = \|\mathbf{P}_i(t) - \mathbf{T}_k\|_2$, and the distance from the Rx $\mathbf{R}_l$ to the target is $R_{\mathrm{rx},i} = \|\mathbf{P}_i(t) - \mathbf{R}_l\|_2$.\par
If the path passes through the wall, the additional optical path difference introduced by the wall is $\Delta R_{\mathrm{wall}} = d_{\mathrm{wall}}(\sqrt{\epsilon_r} - 1)$. Then the total equivalent electrical length is:
\begin{equation}
R_{\Sigma, i}(t) = R_{\mathrm{tx},i} + R_{\mathrm{rx},i} + 2 \cdot \Delta R_{\mathrm{wall}}.
\end{equation}\par
The wall transmission coefficient takes into account the Fresnel transmission coefficient and the dielectric loss. Assuming the free-space impedance $\eta_0 \approx 377~\Omega$ and the wall impedance $\eta_w = \eta_0 / \sqrt{\epsilon_r}$, then the two-way power transmission coefficient is:
\begin{equation}
T_{\mathrm{power}} = \left( \frac{2\eta_w}{\eta_w + \eta_0} \right)^2 \cdot \left( \frac{2\eta_0}{\eta_w + \eta_0} \right)^2.
\end{equation}\par
The wall loss component and the medium attenuation constant are therefore respectively denoted as:
\begin{equation}
L_{\mathrm{wall}} = \exp(-2 \alpha d_{\mathrm{wall}}), \quad \alpha = \frac{\pi f_c}{c} \sqrt{\epsilon_r} \tan\delta.
\end{equation}\par
The echo delay of the $i$-th scattering point is $\tau_i(t) = \frac{R_{\Sigma, i}(t)}{c}$, and the received signal $s_\mathrm{rx}(t_f, t)$ is the superposition of the echoes from all $N_s$ scattering points:
\begin{equation}
s_\mathrm{rx}(t_f, t) = \sum_{i=1}^{N_s} \sigma_i A'_i(t) \cdot s_\mathrm{tx}(t_f - \tau_i(t)),
\end{equation}
where $\sigma_i$ is the radar cross section of the $i$-th scattering point, $A'_i(t)$ includes both free-space path loss and wall attenuation.\par
Perform pulse compression on the received signal. Multiply $s_\mathrm{rx}$ by the conjugate of the transmitted signal $s_\mathrm{tx}^*$. Ignoring the amplitude constant, the beat signal $s_{\mathrm{beat}}(t_f, t)$ after mixing is:
\begin{equation}
\begin{aligned}
s_{\mathrm{beat}}(t_f, t) &= s_\mathrm{rx}(t_f, t) \cdot s_\mathrm{tx}^*(t_f) \\
&= \sum_{i=1}^{N_s} \sigma_i \exp\left( j2\pi \left[ f_c(t_f - \tau_i) + \frac{K(t_f - \tau_i)^2}{2} \right] \right) \\&\quad \quad \quad~\cdot \exp\left( -j2\pi \left[ f_c t_f + \frac{1}{2}K t_f^2 \right] \right)\\
&=\sum_{i=1}^{N_s} \sigma_i \exp\left(j2\pi \left( -f_c \tau_i - K t_f \tau_i + \frac{1}{2}K \tau_i^2 \right)\right)
\end{aligned}.
\end{equation}\par
Neglecting the extremely small residual phase term $\frac{1}{2}K \tau_i^2$, the final baseband signal model is obtained:
\begin{equation}
\begin{aligned}
s_{\mathrm{IF}}(t_f, t) &= \sum_{i=1}^{N_s} \sigma_i \exp\left( -j 2\pi f_c \tau_i(t) \right) \\&\quad\quad\quad~\cdot \exp\left( -j 2\pi K \tau_i(t) t_f \right)
\end{aligned}.
\end{equation}\par

\subsection{RTM and DTM Generation}
Let $n = 0, \dots, N_{\mathrm{ADC}}-1$ be the fast-time sampling index and $m = 0, \dots, N_{\mathrm{pulses}}-1$ be the slow-time sampling index. Discretize the baseband echo signal to obtain:
\begin{equation}
S_{\mathrm{Raw}}[n, m] = s_{\mathrm{IF}}(n/f_s, m \cdot \mathrm{PRT}),
\end{equation}
where $\mathrm{PRT}$ is the pulse repetition time, and $f_s$ is the sampling frequency.\par
To eliminate the clutter introduced by the wall and the static targets, the moving target indication (MTI) filtering is implemented using a two-pulse canceller \cite{RDTM}:
\begin{equation}
S_{\mathrm{MTI}}[n, m] = S_{\mathrm{Raw}}[n, m+1] - S_{\mathrm{Raw}}[n, m].
\end{equation}\par
Perform fast Fourier transform (FFT) on $S_{\mathrm{MTI}}$ along the fast-time dimension $n$ to obtain the complex profile in the range dimension:
\begin{equation}
H(k_r, m) = \sum_{n=0}^{N_{\mathrm{FFT}}-1} S_{\mathrm{MTI}}[n, m] \cdot 
\exp\left(-j \frac{2\pi}{N_{\mathrm{FFT}}} n k_r\right),
\end{equation}
where $N_{\mathrm{FFT}}$ is the number of FFT points, and $k_r$ is the range bin index. The corresponding physical range $r$ is:
\begin{equation}
r = \frac{c f_s}{2K} \cdot \frac{k_r}{N_{\mathrm{FFT}}}.
\end{equation}\par
The RTM is obtained after taking the modulus:
\begin{equation}
\mathbf{I}_{\mathrm{RTM}}(r, t_m) = \left| H(k_r, m) \right|,
\end{equation}
where $t_m$ is the discrete slow-time index.
Calculate the distance $R_{\mathrm{torso}}(t_m)$ of the torso relative to the radar center at each moment $t_m$:
\begin{equation}
R_{\mathrm{torso}}(t_m) = \|\mathbf{P}_{\mathrm{Torso}}(t_m) - \mathbf{P}_{\mathrm{RadarCenter}}\|_2.
\end{equation}\par
To offset the phase change $\exp(-j \frac{4\pi}{\lambda} R_{\mathrm{torso}})$ caused by torso motion, a conjugate compensation term is constructed:
\begin{equation}
C_{\mathrm{comp}}(m) = \exp\left( j \frac{4\pi}{\lambda} R_{\mathrm{torso}}(m) \right),
\end{equation}
where $\lambda$ is the wave length. sum the compressed complex signal $H(k_r, m)$ along the range dimension to obtain a slow-time signal, and optionally multiply it by the compensation term:
\begin{equation}
\begin{aligned}
S_{\mathrm{sum}}(m) &= \sum_{k_r} H(k_r, m)\\
S_{\mathrm{comp}}(m) &= S_{\mathrm{sum}}(m) \cdot C_{\mathrm{comp}}(m)
\end{aligned}.
\end{equation}\par
This step may shift the center of the Doppler spectrum from $f_D = \frac{2v_{\mathrm{radial}}}{\lambda}$ to $0~\mathrm{Hz}$, so that the micro-Doppler signature of the limb are distributed on both sides of the zero frequency. Perform STFT analysis on the compensated signal $S_{\mathrm{comp}}(m)$ using a Hamming window $w[\eta]$:
\begin{equation}
D(f_d, m) = \sum_{\eta = -\frac{L_w}{2}}^{\frac{L_w}{2}-1} S_{\mathrm{comp}}[m + \eta] \cdot w[\eta] \cdot e^{-j 2\pi f_d \eta},
\end{equation}
where $f_d$ is the Doppler frequency index.\par
The DTM is obtained after taking the modulus \cite{RDTM}:
\begin{equation}
\mathbf{I}_{\mathrm{DTM}}(f_d, t_m) = \left| D(f_d, m) \right|.
\end{equation}\par
\begin{figure*}
    \centering
    \includegraphics[width=\textwidth]{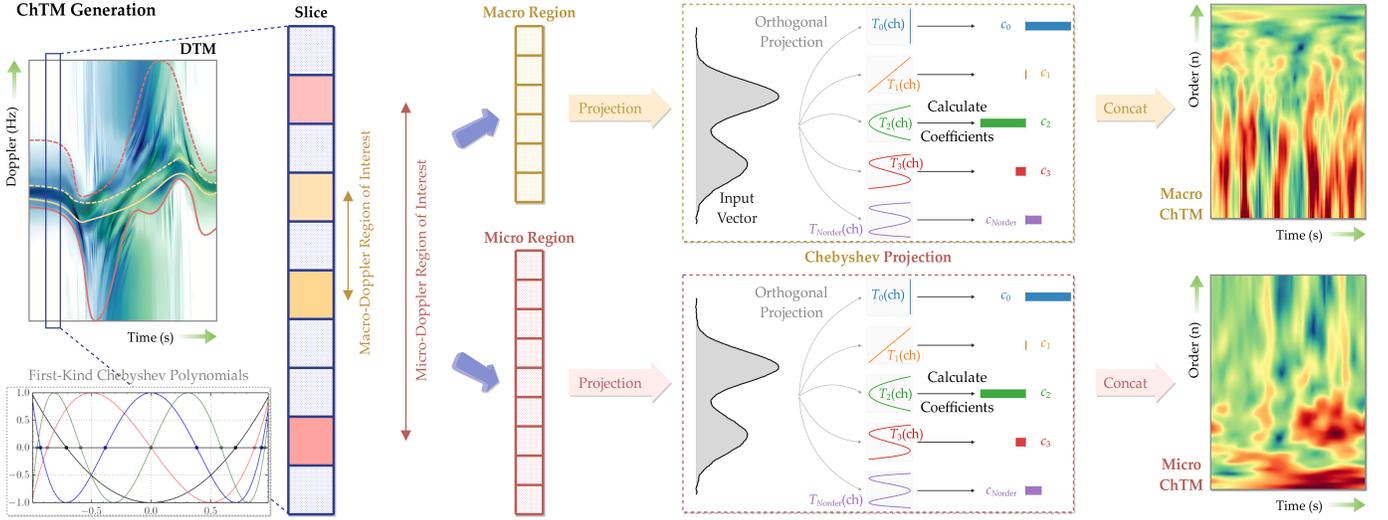}
    \caption{Schematic diagram of the proposed DTM envelope extraction and ChTM generation method.}
    \label{ChTM_Generation}
    \vspace{-0.2cm}
\end{figure*}\par

\section{ChTM Generation and Analysis}
In this section, the definitions of macro and micro micro-Doppler envelopes are first presented, and a method for extracting the envelopes is proposed. Secondly, the method of projecting DTM to ChTM is proposed. Finally, the physical background of the micro-Doppler signature representation on the ChTM is discussed.\par

\subsection{Macro and Micro-Doppler Envelopes}
The macro and micro-Doppler envelopes are important features for characterizing the boundary of human body motion \cite{Envelope}. The definitions of the four envelopes are as follows: (1) The upper/lower micro-Doppler envelopes are used to characterize the fastest motion speed of the limb ends. (2) The upper/lower macro Doppler envelopes are used to characterize the main energy distribution range of the human torso centroid.\par
The DTM matrix generated in the previous section is $\mathbf{I}_{\mathrm{DTM}} \in \mathbb{R}^{N_f \times N_t}$, where $N_f$ is the number of Doppler frequency sampling points and $N_t$ is the number of slow-time sampling points. Define $I(k, m)$ as the normalized amplitude value at the $k$-th frequency index $f_k$ and the $m$-th time index $t_m$, where $k \in [1, N_f]$ and $m \in [1, N_t]$. In order to suppress the speckle noise in the TWR echo and preserve the edge features, a two- dimensional Gaussian smoothing process is first applied to the DTM. Let the Gaussian kernel be $G(u, v)$ with a standard deviation of $\sigma_g$:
\begin{equation}
\begin{aligned}
I_{\mathrm{smooth}}(k, m) &= I(k, m) \ast G(u, v) \\
&= \sum_{u} \sum_{v} I(k-u, m-v) \cdot \frac{\exp\left( -\frac{u^2 + v^2}{2\sigma_g^2} \right)}{2\pi \sigma_g^2}
\end{aligned},
\end{equation}
where $\sigma^2_g$ is the variance of the Gaussian kernel.\par
To distinguish between high-energy torso echoes and low-energy limb echoes, two adaptive thresholds are defined. For micro-Doppler signature, define $\eta_{\mathrm{mD}} = \lambda_{\mathrm{mD}} \cdot \max_{k,m} \{ I_{\mathrm{smooth}}(k, m) \}$; for macro-Doppler signature, define $\eta_{\mathrm{torso}} = \lambda_{\mathrm{torso}} \cdot \max_{k,m} \{ I_{\mathrm{smooth}}(k, m) \}$, where $\lambda_{\mathrm{mD}},\lambda_{\mathrm{torso}}$ are the ratio to the max micro-Doppler and macro-Doppler amplitude for adaptively selecting the threshold, respectively.\par
For each moment $m$, the frequency axis $k$ is scanned and the discrete indices of the four envelopes are calculated. First, the upper micro-Doppler envelope $E_{\mathrm{mD}}^{\mathrm{up}}(m)$ corresponds to the maximum forward velocity of the limb. The frequency indices can be calculated as:
\begin{equation}
k_{\mathrm{up}}^{\mathrm{mD}}(m) = \min \{ k \mid I_{\mathrm{smooth}}(k, m) > \eta_{\mathrm{mD}} \}.
\end{equation}\par
Second, the lower micro-Doppler envelope $E_{\mathrm{mD}}^{\mathrm{low}}(m)$ corresponds to the maximum backward velocity of the limb. The frequency indices can be calculated as:
\begin{equation}
k_{\mathrm{low}}^{\mathrm{mD}}(m) = \max \{ k \mid I_{\mathrm{smooth}}(k, m) > \eta_{\mathrm{mD}} \}.
\end{equation}\par
Third, the upper macro-Doppler envelope $E_{\mathrm{torso}}^{\mathrm{up}}(m)$ corresponds to upper boundary of the torso Doppler spectrum. The frequency indices can be calculated as:
\begin{equation}
k_{\mathrm{up}}^{\mathrm{torso}}(m) = \min \{ k \mid I_{\mathrm{smooth}}(k, m) > \eta_{\mathrm{torso}} \}.
\end{equation}\par
Finally, the lower macro-Doppler envelope $E_{\mathrm{torso}}^{\mathrm{up}}(m)$ corresponds to lower boundary of the torso Doppler spectrum. The frequency indices can be calculated as:
\begin{equation}
k_{\mathrm{low}}^{\mathrm{torso}}(m) = \max \{ k \mid I_{\mathrm{smooth}}(k, m) > \eta_{\mathrm{torso}} \}.
\end{equation}\par
In particular, if no valid indices are detected at a certain moment, filling is performed by linear interpolation. Finally, moving median filtering and locally estimated scatterplot smoothing are used to obtain a continuous envelope $f_\mathrm{env}(t)$. The four continuous envelopes will be used for subsequent ChTM generation.\par
In the following, based on the human kinematic model established in the previous text, the physical velocity limits corresponding to the extracted envelopes are derived.\par
Suppose the phase compensation has been performed before generating the DTM. Therefore, theoretically, the Doppler frequency at the center of the torso is shifted to $0\mathrm{~Hz}$. However, the torso is not an ideal point target but a scatterer with a certain volume. The torso envelope $E_{\mathrm{torso}}(t)$ reflects the small relative velocities of each scattering point on the torso with respect to the center of the torso and the broadening of their radar cross section. Its physical counterpart is the relative swinging velocities of the shoulders and hips:
\begin{equation}
v_{\mathrm{torso\_env}}(t) \approx \frac{\lambda}{2} \cdot \Delta f_{\mathrm{width}},
\end{equation}
where $\Delta f_{\mathrm{width}}$ corresponds to the torso Doppler broadening caused by slight shoulder torsion during normal walking.\par
The micro-Doppler envelope $E_{\mathrm{mD}}(t)$ directly corresponds to the instantaneous Doppler frequency of the joint with the maximum absolute value of the radial velocity among all human joints. For walking, the maximum velocity is usually generated by the ankles or hands. Take the right ankle as an example, as:
\begin{equation}
\mathbf{P}_{\mathrm{RA}}(t) = \mathbf{P}_{\mathrm{RK}}(t) + \mathbf{R}_y(\theta_{\mathrm{ra}}(t)) \cdot [0, 0, -L_{\mathrm{calf}}]^\top,
\end{equation}
where $\theta_{\mathrm{ra}}(t) = A_{\mathrm{calf}}\sin(2\pi f_{\mathrm{gait}} t + \frac{\pi}{4})$. Known the radar is located in the front along the $x^{+}$ direction, and the person walks along the $x$-axis. The limb swing mainly occurs in the $xOz$ plane. Focus on the radial velocity $v_r$ in the $x$-axis direction. Ignore the influence of the slight change in the height $z$ on the radial direction, and only consider the derivative of the displacement along the $x$-axis.\par
The hip speed is a constant $v_{\mathrm{torso}}$. The position offset of the right knee relative to the hip along the $x$-axis is:
\begin{equation}
x_{\mathrm{knee/hip}}(t) = -L_{\mathrm{thigh}} \cdot \sin(\theta_{\mathrm{rk}}(t)).
\end{equation}\par
The position offset of the right ankle relative to the right knee along the $x$-axis is:
\begin{equation}
x_{\mathrm{ankle/knee}}(t) = -L_{\mathrm{calf}} \cdot \sin(\theta_{\mathrm{ra}}(t)).
\end{equation}\par
Taking the derivative with respect to time $t$, the instantaneous radial velocity $v_{\mathrm{RA}}(t)$ of the right ankle is obtained:
\begin{equation}
\begin{aligned}
v_{\mathrm{RA}}(t) &= v_{\mathrm{torso}} - L_{\mathrm{thigh}} \dot{\theta}_{\mathrm{rk}}(t) \cos(\theta_{\mathrm{rk}}(t)) \\&- L_{\mathrm{calf}} \dot{\theta}_{\mathrm{ra}}(t) \cos(\theta_{\mathrm{ra}}(t))
\end{aligned}.
\end{equation}\par
Known $\theta_{\mathrm{rk}}(t) = A_{\mathrm{thigh}}\sin(2\pi f_{\mathrm{gait}} t)$, then the angular velocity $\dot{\theta}_{\mathrm{rk}}(t) = 2\pi f_{\mathrm{gait}} A_{\mathrm{thigh}} \cos(2\pi f_{\mathrm{gait}} t)$. Therefore, the physical frequency value $f_{\mathrm{mD}}(t)$ of the micro-Doppler envelope at time $t$ satisfies:
\begin{equation}
f_{\mathrm{mD}}(t) = \frac{2}{\lambda} \cdot \max_{i \in \text{All Joints}} \left| v_{r, i}(t) - v_{\mathrm{comp}}(t) \right|.
\end{equation}\par
Due to torso compensation, $v_{\mathrm{comp}} \approx v_{\mathrm{torso}}$, thus the envelope mainly reflects the swinging velocity term of the limb relative to the torso:
\begin{equation}
f_{\mathrm{env}}(t) \propto \frac{2}{\lambda} \left( L_{\mathrm{thigh}} \omega_{\mathrm{gait}} A_{\mathrm{thigh}} + L_{\mathrm{calf}} \omega_{\mathrm{gait}} A_{\mathrm{calf}} \right).
\end{equation}\par
This indicates that the amplitude of the micro-Doppler envelope is proportional to the gait frequency $f_{\mathrm{gait}}$, limb length, and swing amplitude. If phase compensation is not performed when generating the DTM, the micro-Doppler signature will shift on the frequency axis, but the derived conclusions will remain unchanged.\par

\subsection{DTM to ChTM Projection}
To characterize the time-frequency texture features inside the envelope, the Chebyshev polynomial is introduced to perform approximate orthogonal decomposition on the local slice of the DTM \cite{Chebyshev}, thereby generating two images: the macro ChTM based on the torso envelope and the micro ChTM based on the global micro-Doppler envelope.\par
For each moment $m$, determine the effective range of the Doppler frequency $[k_{\mathrm{start}}, k_{\mathrm{end}}]$ based on the extracted envelope. The macro region of interest is $k \in [k_{\mathrm{up}}^{\mathrm{torso}}(m), k_{\mathrm{low}}^{\mathrm{torso}}(m)]$. The micro region of interest is $k \in [k_{\mathrm{up}}^{\mathrm{mD}}(m), k_{\mathrm{low}}^{\mathrm{mD}}(m)]$.\par
Extract the spectrum slice vector $\mathbf{s}_m \in \mathbb{R}^{N_s}$ within this range, where $N_s = k_{\mathrm{end}} - k_{\mathrm{start}} + 1$. In order to utilize the orthogonality of Chebyshev polynomials, the discrete frequency index $j \in [1, N_s]$ must be mapped to the standard domain $\mathrm{ch} \in [-1, 1]$:
\begin{equation}
\mathrm{ch}_j = -1 + \frac{2(j - 1)}{N_s - 1}, \quad j = 1, 2, \dots, N_s.
\end{equation}\par
If $N_s < 2$, set the coefficient to zero at this moment.\par
Decompose the signal $\mathbf{s}_m$ using the first-kind Chebyshev polynomials $T_n(\mathrm{ch})$ as the basis functions. The Chebyshev polynomials are defined by the recurrence formula:
\begin{equation}
\begin{cases}
T_0(\mathrm{ch}) = 1 \\
T_1(\mathrm{ch}) = \mathrm{ch} \\
T_n(\mathrm{ch}) = 2\mathrm{ch}\cdot T_{n-1}(\mathrm{ch}) - T_{n-2}(\mathrm{ch}), \quad n \ge 2
\end{cases}.
\end{equation}\par
Let the maximum decomposition order to be $N_{\mathrm{order}}$. For each order $n \in [0, N_{\mathrm{order}}]$, calculate the projection coefficient $c_{n,m}$ of the signal $\mathbf{s}_m$ on the basis function $T_n(x)$. Use the matrix operation form. Construct the basis function matrix $\mathbf{T} \in \mathbb{R}^{N_s \times (N_{\mathrm{order}}+1)}$, where the element in the $j$-th row and the $n$-th column is $T_n(\mathrm{ch}_j)$. The coefficient vector $\mathbf{c}_m = [c_{0,m}, c_{1,m}, \dots, c_{N_{\mathrm{order}},m}]^\top$ is calculated as follows:
\begin{equation}
\mathbf{c}_m = \frac{1}{N_s} \mathbf{T}^\top \cdot \mathbf{s}_m,
\end{equation}
where $N_s$ is served as the normalization factor.\par
Stack the coefficient vectors at all time instants, take the logarithmic magnitude, and obtain the ChTM:
\begin{equation}
\mathbf{I}_{\mathrm{ChTM}}(n, m) = \log_{10}\left( \mathrm{Norm}(c_{n,m}) + \epsilon \right),
\end{equation}
where $\epsilon$ is a very small bias, and $\mathrm{Norm}(\cdot)$ is the normalization function processed over the whole image.\par
In the following, the physical background of ChTM is discussed in detail. Only micro-Doppler envelope based ChTM is used for discussion. The macro-Doppler envelope based ChTM can be used in some recognition scenarios for information compensation.\par
Although $c_{n,m}$ is obtained through discrete matrix operations, in theoretical analysis, it can be regarded as the orthogonal projection of the DTM slice onto the Chebyshev basis functions. For time $m$, the continuous spectrum function in the normalized frequency domain $\mathrm{ch} \in [-1, 1]$ is denoted as $S_m(\mathrm{ch})$. According to the orthogonality of Chebyshev polynomials, the coefficient $c_{n,m}$ can be approximately expressed in the form of a weighted integral:
\begin{equation}
c_{n,m} \propto \int_{-1}^{1} S_m(\mathrm{ch}) T_n(\mathrm{ch}) \frac{1}{\sqrt{1-\mathrm{ch}^2}} d\mathrm{ch},
\end{equation}
where the weight function $(1-\mathrm{ch}^2)^{-1/2}$ assigns higher weights to the envelope, which is highly consistent with the physical characteristic that micro-Doppler signature tend to be concentrated at the envelope edges.\par
When $n = 0$, the basis function $T_0(\mathrm{ch}) = 1$. Then the discrete representation of the coefficient $c_{0,m}$ is:
\begin{equation} 
c_{0,m} = \frac{1}{N_s} \sum_{j=1}^{N_s} \mathbf{s}_m[j] \cdot 1 = \frac{1}{N_s} \sum{j=1}^{N_s} \mathbf{s}_m[j],
\end{equation}
which means that $c_{0,m}$ characterizes the average energy intensity of the echo signal in the region of interest at that moment. For the Macro ChTM, $c_{0,m}$ reflects the RCS scintillation characteristics of the torso caused by the change of the azimuth angle. For the Micro ChTM, $c_{0,m}$ reflects the total echo energy of the scattering points of the whole human body. When the RCS increases due to the strenuous movement of the human body, this coefficient increases significantly.\par
When $n = 1$, the basis function $T_1(\mathrm{ch}) = \mathrm{ch}$. Then the discrete representation of the coefficient $c_{1,m}$ is:
\begin{equation} 
c_{1,m} = \frac{1}{N_s} \sum_{j=1}^{N_s} \mathbf{s}_m[j] \cdot \mathrm{ch}_j. 
\end{equation}\par
This equation has the form of the first-order moment, so $c_{1,m}$ represents the linear slope or centroid offset of the DTM energy relative to the geometric center. Since $\mathrm{ch} \in [-1, 1]$ corresponds to the Doppler range from the upper envelope to the lower envelope, if $c_{1,m} > 0$, it indicates that the spectrum energy is mainly concentrated at the positive frequency end, which physically corresponds to the accelerating motion of the limb towards the TWR. If $c_{1,m} < 0$, it indicates that the spectrum energy is concentrated at the negative frequency end, which physically corresponds to the motion of the limb away from the TWR. Therefore, the sign and amplitude of $c_{1,m}$ can sensitively capture the asymmetry of human motion in the positive and negative Doppler directions.\par
When $n = 2$, the basis function $T_2(\mathrm{ch}) = 2\mathrm{ch}^2 - 1$. The coefficient $c_{2,m}$ is expressed as:
\begin{equation}
c_{2,m} = \frac{1}{N_s} \sum_{j = 1}^{N_s} \mathbf{s}_m[j] (2\mathrm{ch}_j^2 - 1)= \frac{2}{N_s} \sum_{j = 1}^{N_s} \mathbf{s}_m[j] \mathrm{ch}_j^2 - c_{0,m}.
\label{c2m}
\end{equation}
The first term contains $\mathbf{s}_m[j] \mathrm{ch}_j^2$, which has the form of the second-order moment and describes the dispersion degree of the energy distribution. Then $c_{2,m}$ characterizes the broadening of the DTM and the concavity-convexity of its shape. For normal walking, the limbs of the human body swing with large amplitudes and the speed distribution range is wide. The DTM has energy distribution in the entire region of interest, resulting in a relatively large $\mathrm{ch}_j^2$ and keeping $c_{2,m}$ at a relatively high level. For armed walking, the upper limbs of the human body are locked, only the lower limbs swing, and the overall movement is rigid. The Doppler energy is highly concentrated around the trunk speed, showing relatively narrow-band characteristics. At this time, $\mathrm{ch}_j^2 \approx 0$, causing $c_{2,m}$ to decrease significantly or even become negative. Therefore, the second-order coefficient is theoretically a key low-order feature for distinguishing between normal walking and armed walking.\par
When $n$ is relatively large, the Chebyshev polynomial $T_n(\mathrm{ch})$ exhibits high-frequency oscillation characteristics within the interval. Then the high-order coefficients characterize the fine texture and structural complexity inside the DTM. According to the kinematic model established above:
\begin{equation}
s_{\mathrm{IF}}(t_f, t) = \sum_{i} \sigma_i \cdot \exp\left(-j 2\pi f_c \tau_i(t)\right).
\end{equation}\par
The DTM is the coherent superposition of the echoes from multiple limb joints. Therefore, during normal walking, the non-rigid motion of multiple limbs leads to the appearance of multiple local peaks and complex interference fringes in the spectrum. These high-frequency fluctuations require high-order Chebyshev polynomials to fit, so the high-order coefficients have significant energy. During armed walking, the arms are relatively stationary with respect to the torso, the relative motion between scattering points is reduced, and the inside of the envelope is relatively smooth. At this time, the high-order coefficients show a decaying trend as $n$ increases.\par
In summary, ChTM projects the high-dimensional DTM to a coefficient space with clear physical meanings. The low-order coefficients $n = 0, 1, 2$ lock in the overall motion trend and spectral width of the target, while the high-order coefficients $n>2$ record the fine structure of limb micro-motions, thus providing a robust feature representation for distinguishing similar human activities. Moreover, since the number of its orders is often smaller than the number of frequency points of DTM, it can effectively compress the quantity of parameters.\par

\subsection{Theoretical Proof of the Effectiveness of ChTM}
In this section, a more rigid theoretical derivation of the effectiveness of ChTM in characterizing micro-Doppler signature will be carried out from the perspectives of information theory and function approximation theory \cite{Function_Approximation}. The main objective is to prove that, under a finite Chebyshev order $N_{\mathrm{order}}$, the amount of information $I_{\mathrm{ChTM}}$ in the feature space retained by ChTM is sufficient to cover the minimum amount of kinematic difference information $I_{\mathrm{req}}$ required to distinguish normal walking from armed, that is, to prove $I_{\mathrm{ChTM}} \ge I_{\mathrm{req}}$.\par
According to the established human kinematic model, the radar echo of a human target is the coherent superposition of the echoes from each scattering point. For the radar line-of-sight direction, the set of the radial velocity distribution of the human body at time $t$ is defined as $\mathcal{V}(t)$. For normal walking, both the upper and lower limbs are involved in the swinging motion, and their velocity sets can be expressed as:
\begin{equation}
\mathcal{V}_{\mathrm{NW}}(t) = \{ v_{\mathrm{torso}} \} \cup \{ v_{\mathrm{leg}, i}(t) \}_{i=1}^4 \cup \{ v_{\mathrm{arm}, j}(t) \}_{j=1}^4,
\end{equation}
where $v_{\mathrm{leg}}$ and $v_{\mathrm{arm}}$ respectively include the radial velocities of the thighs, calves, upper arms, and forearms. According to Eq. (\ref{Normal_RE}) and (\ref{Normal_RH}), the velocities of the arm joints contain an alternating current component that varies over time, that is, $|v_{\mathrm{arm}, j}(t) - v_{\mathrm{torso}}| > 0$.\par
For armed walking, the upper limbs are locked, and its velocity set degenerates to:
\begin{equation}
\mathcal{V}_{\mathrm{AW}}(t) = \{ v_{\mathrm{torso}} \} \cup \{ v_{\mathrm{leg}, i}(t) \}_{i=1}^4 \cup \{ v_{\mathrm{gun}} \approx v_{\mathrm{torso}} \},
\end{equation}
where the upper limbs and the torso remain relatively stationary, and the speed difference approaches zero.\par
To achieve effective recognition, the feature space must be able to distinguish the differences between these two types of motion sets. Define the instantaneous difference function $D_v(t)$ of the two types of activities in the velocity domain as the Hausdorff distance between the two sets:
\begin{equation}
D_v(t) = \max_{v_a \in \mathcal{V}_{\mathrm{NW}}} \min_{v_b \in \mathcal{V}_{\mathrm{AW}}} |v_a(t) - v_b(t)|.
\end{equation}\par
Since the lower limb motion is basically the same in the two modes, the differences are mainly contributed by the upper limb swings. Substituting the angular velocity $\dot{\theta}_{\mathrm{re}}(t)$ in Eq. (\ref{Normal_RE}), the lower bound of the difference can be obtained:
\begin{equation}
D_v(t) \approx |v_{\mathrm{arm}}(t) - v_{\mathrm{torso}}| \propto 2\pi f_{\mathrm{gait}} A_{\mathrm{arm}} L_{\mathrm{arm}}.
\end{equation}\par
This indicates that the amount of information $I_{\mathrm{req}}$ required for effective recognition is essentially the ability to reconstruct the micro-Doppler modulation component with a frequency of $f_{\mathrm{gait}}$ and an amplitude proportional to $A_{\mathrm{arm}}$.\par
On the other hand, based on the TWR echo model, $\mathbf{s}_m$ is the mapping of the velocity distribution $\mathcal{V}(t)$ in the frequency domain. As the Doppler frequency corresponding to the velocity $v$ be $f_d = 2v/\lambda$. Ideally, the DTM slice $S_m(f)$ can be modeled as the superposition of multiple point spread functions of scattering centers:
\begin{equation}
S_m(f) = \sum_{k \in \text{Scatterers}} \sigma_k \cdot \Psi(f - \frac{2v_k(t_m)}{\lambda}),
\end{equation}
where $\Psi(\cdot)$ represents the shape of the main lobe of the spectrum determined by the window function.\par
During normal walking, due to the existence of the $v_{\mathrm{arm}}$ component, $S_m(f)$ exhibits multiple separated local peaks within the micro-Doppler frequency band, and the energy distribution is relatively dispersed. During armed walking, due to the absence of $v_{\mathrm{arm}}$ or its incorporation into $v_{\mathrm{torso}}$, the number of peaks of $S_m(f)$ decreases, and the energy is more concentrated near the compensated torso velocity, showing a single-peak or few-peak pattern. Therefore, the recognition problem is transformed into: If finite-order Chebyshev coefficients can be used to distinguish whether $S_m(f)$ has a multi-peak broadband shape or a single-peak narrowband shape.\par
On the normalized frequency domain $\mathrm{ch} \in [-1, 1]$, the DTM slicing function $S_m(\mathrm{ch})$ is a square-integrable function. According to the Chebyshev series theory \cite{Chebyshev_Theory}, its $N_{\mathrm{order}}$-th order truncation error $E_{N_{\mathrm{order}}}$ satisfies:
\begin{equation}
E_{N_{\mathrm{order}}} = \| S_m(\mathrm{ch}) - \sum_{n=0}^{N_{\mathrm{order}}} c_{n,m} T_n(\mathrm{ch}) \|_2^2 = \sum_{n=N_{\mathrm{order}}+1}^{\infty} c_{n,m}^2.
\end{equation}\par
Since the continuous Doppler spectrum generated by human joint motion is a smooth function, the Chebyshev coefficients $c_{n,m}$ decay exponentially with the order $n$.
The topological structures such as the number and positions of the spectral peaks are the valid information that needs to be retained. According to the polynomial interpolation theory \cite{Polynomial_Interpolation}, an $n$-th order polynomial can describe at most $n - 1$ extreme points. For the human micro-Doppler signature, the main energy contributors are: the torso ($1$), the left/right arms ($\ge 2$), and the left/right legs ($\ge 2$). At the most complex moment, there may be about $5$ main peaks appearing simultaneously on the spectral slice $S_m(\mathrm{ch})$. In order to retain these peak features in the fitted curve, the minimum required order $N_{\mathrm{req}}$ should satisfy:
\begin{equation}
N_{\mathrm{req}} \ge 2 \times (\text{Number of Peaks}) \approx 10,
\end{equation}
which means that as long as the set decomposition order $N_{\mathrm{order}} \ge N_{\mathrm{req}}$, the truncated Chebyshev series can recover the multi-peak structure of the spectrum in the topological sense.\par
Further analyze the sensitivity of the coefficients $c_{n,m}$ of different orders to the difference $D_v(t)$. Based on the discussion from the last subsection, when $n = 2$, according to Eq. (\ref{c2m}), $c_{2,m}$ corresponds to the second-order moment of the spectrum. Due to the existence of $v_{\mathrm{arm}}$ during normal walking, the spectral width is significantly larger than that during armed walking. Therefore, $|c_{2,m}^{\mathrm{NW}}| \gg |c_{2,m}^{\mathrm{AW}}|$. This is the most prominent feature for distinguishing between the two types of targets. When $n > 2$, the high-order coefficients encode the detailed texture of the spectrum. For armed walking, since the arms are stationary, the spectrum is smooth, and the high-order coefficients $c_{n,m}^{\mathrm{AW}}$ rapidly decay to the noise level. In contrast, for normal walking, due to the interference of slight motion of multiple limbs, there are high-frequency oscillations in the spectrum, and the high-order coefficients $c_{n,m}^{\mathrm{NW}}$ still maintain a relatively high amplitude.\par
Define the feature vector extracted by ChTM at time $m$ as $\mathbf{c}_m \in \mathbb{R}^{N_{\mathrm{order}}+1}$. The distance $D_{\mathrm{ChTM}}$ between the two types of activities in the feature space is:
\begin{equation}
D_{\mathrm{ChTM}}(m) = \| \mathbf{c}_m^{\mathrm{NW}} - \mathbf{c}_m^{\mathrm{AW}} \|_2.
\end{equation}\par
According to the form of Parseval's theorem under the orthogonal basis \cite{Parseval}, ignoring the truncation error, the Euclidean distance in the feature space is equivalent to the distance of the original function in the weighted $L^2$ space:
\begin{equation}
D_{\mathrm{ChTM}}(m) \approx \| S_m^{\mathrm{NW}}(\mathrm{ch}) - S_m^{\mathrm{AW}}(\mathrm{ch}) \|_{wL^2},
\end{equation}
where $S_m^{\mathrm{NW}}(\mathrm{ch})$ and $S_m^{\mathrm{AW}}(\mathrm{ch})$ are DTM slices of normal walking and armed walking, respectively. Due to the significant physical difference between $S_m^{\mathrm{NW}}$ and $S_m^{\mathrm{AW}}$, the spectral distributions do not overlap with each other. Therefore, $D_{\mathrm{ChTM}}(m)$ is significantly greater than zero. Therefore, as long as an appropriate $N_{\mathrm{order}}$ ($10\ll N_{\mathrm{order}}< N_f$) is selected, the ChTM feature vectors can comprehensively represent the kinematic difference information required to distinguish between the two types of activities. This theoretically proves that the ChTM does not lose the key classification features while reducing the dimensionality, which is a physically interpretable and compressed micro-Doppler representation.$\hfill \square$\par
\begin{table}[!t]
\begin{center}
\caption{Experimental Parameters and Settings$^{*}$.\label{Experimental Parameters}}
\vspace{-0.1cm}
\resizebox{0.48\textwidth}{!}{
\begin{tabular}{cc}
\hline\hline
\textbf{Parameters}             & \textbf{Value}     \\ 
\hline
\multicolumn{2}{c}{\textbf{Parameters of Simulated and Measured TWR System}}    \\
\hline
Settings of MIMO Antenna     &  $8$ Tx, $8$ Rx, $64$ Channels  \\
Antenna Center Height to Ground & $1.5 \mathrm{~m}$      \\ 
Center Frequency          & $2.5 \mathrm{~GHz}$   \\
Bandwidth & $1.0 \mathrm{~GHz}$  \\
Fast-Time Sampling Points & $3190$                \\
Slow-Time Sampling Rate$^{1}$ & $200/s$             \\
Sampling Period & $1 \mathrm{~s}$                 \\
\hline
\multicolumn{2}{c}{\textbf{Parameters of the Scenario}}        \\
\hline
Wall Thickness & $0.24 \mathrm{~m}$  \\
Wall Relative Dielectric Constant & $6$ (Estimated) \\
Human Motion Range from Radar & $1 \sim 5 \mathrm{~m}$     \\
Classes of Tester Identities$^{2}$ & $4$ \\
Classes of Threat$^{3}$ & $2$ \\
Training Set$^{4}$ & $1197$ \\
Validation Set$^{4}$ & $299$ \\
\hline\hline
\end{tabular}
}
\end{center}
\footnotesize $^{*}$ Simulated and measured datasets are conducted under the same parameters.\\
\footnotesize $^{1}$ The PRF of the system is $1000$. A $5$-fold downsampling is employed during system data acquisition.\\
\footnotesize $^{2}$ $4$ different testers are included in the experiment.\\
\footnotesize $^{3}$ Unarmed and armed states are included in the experiment.\\
\footnotesize $^{4}$ The data volume across categories is unbalanced, approximating the complex application conditions.\\
\vspace{-0.2cm}
\end{table}\par

\section{Numerical Simulations and Experiments}
In this section, the generation schemes of simulated and measured datasets and the scenario settings utilized for experiments are first introduced. Furthermore, visualization verification and comparative verification of the proposed method are conducted sequentially. Finally, the experimental results are discussed in detail.\par
\begin{table}
\begin{center}
\caption{Uniform Hyperparameters for Network Models$^{*}$.\label{Training Settings}}
\vspace{-0.1cm}
\resizebox{0.48\textwidth}{!}{
\begin{tabular}{cc}
\hline\hline
\textbf{Name of Hyperparameters}             & \textbf{Value}          \\ \hline
Batch Size                      & $32$                     \\
Total Epoches                   & $60$                    \\
Initial Learning Rate           & $0.00147$                   \\
Regulization Method             & $L - 2$           \\
Optimizer                       & Adam  \\
Solidified Model                & Best Validation Loss                \\
Training Hardware$^{1}$      & NVIDIA Tesla V100 \\
Validation Hardware$^{1}$      & NVIDIA RTX 3060 OC \\
Training / Validation Software  & Python 3.10, Paddlepaddle 3.0.0   \\
\hline\hline
\end{tabular}
}
\end{center}
\footnotesize $^{*}$ The hyperparameters for all used network models remain the same.\\
\footnotesize $^{1}$ The training is completed on the server, but the validation is completed on the local host. The inference speed can only be compared on the local host.
\vspace{-0.2cm}
\end{table}\par

\subsection{Dataset and Scenario Settings}
Simulated and measured TWR HAR datasets are jointly utilized to verify the proposed method, based on the system and scenario parameter settings indicated in TABLE \ref{Experimental Parameters}. The simulated dataset is generated from the TWR echo model, while the measured dataset is gathered in typical urban scenarios using a self-developed ultra-wideband TWR system.\par
During the experiment, linear FMCW waves ranging from $2$ to $3\,\mathrm{GHz}$ are transmitted and received. An $8$-transmit and $8$-receive multiple-input multiple-output antenna array is positioned $1.5\,\mathrm{m}$ above the ground. For the fast-time dimension, $3190$ sampling points are acquired, and $200$ sampling points per second are collected for the slow-time dimension. The fast-time axis corresponding to range bins is truncated from $0$ to $6$ meters, and the slow-time axis corresponding to time windows is truncated from $0$ to $1$ second. A single hollow brick wall with a thickness of $0.24\,\mathrm{m}$ is employed in the experiments. In the simulations, this wall is approximated as a $5 \times 0.24 \times 2.5\,\mathrm{m}$ isotropic rectangular medium, where a relative dielectric constant of $6$ and a loss tangent of $0.03$ are selected. For data acquisition in the measured experiments, a concrete brick wall characterized by approximately the same thickness, a dielectric constant of about $6$, and a loss tangent of about $0.03$ is utilized. Eight categories are designed for the labels in both simulated and measured datasets. The tester identity categories are composed of four different personnel denoted as $\mathrm{P1}$ to $\mathrm{P4}$, and the threat categories are distinguished by unarmed and armed states.\par
\begin{figure*}
    \centering
    \includegraphics[width=\textwidth]{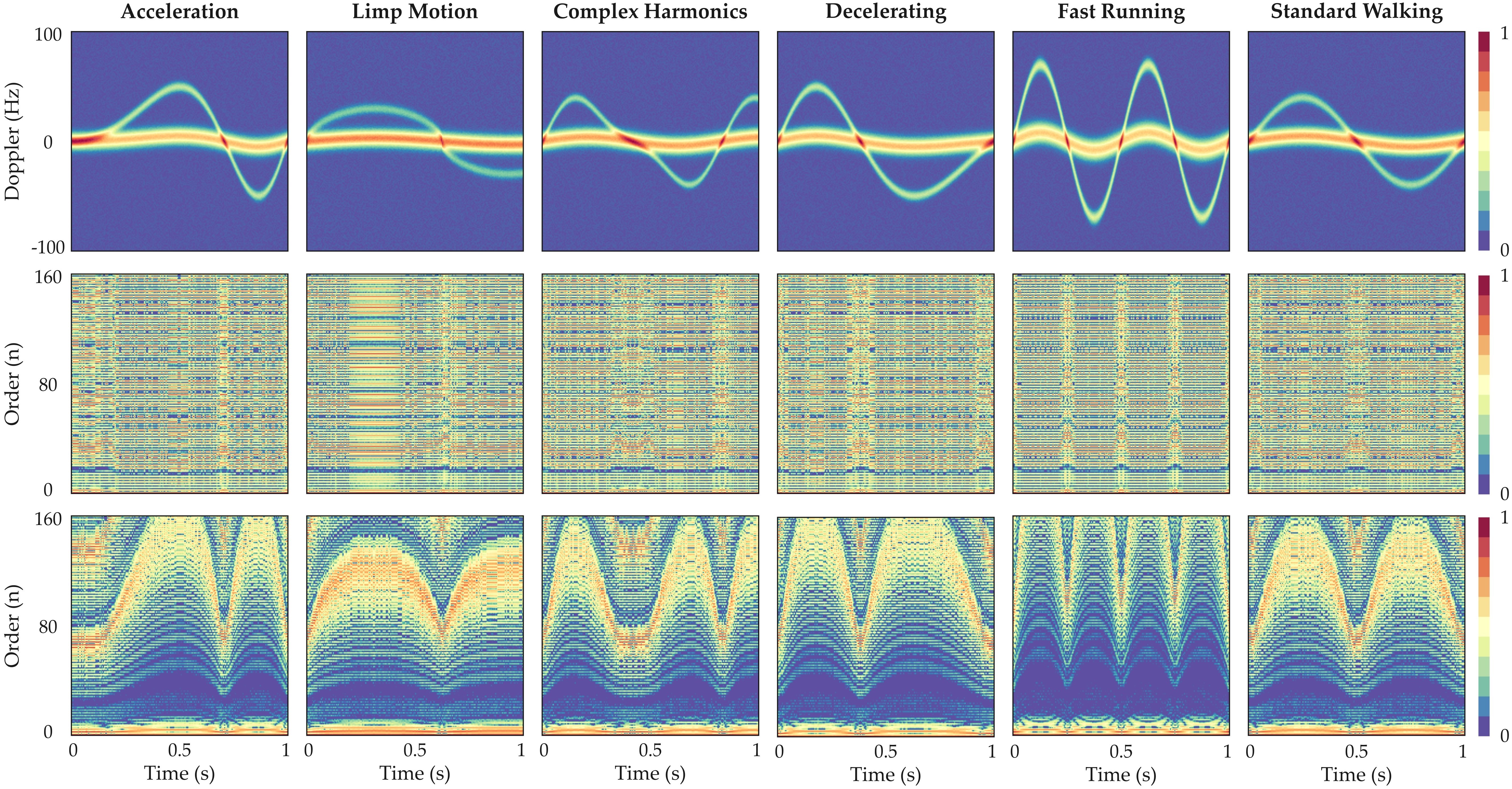}
    \caption{Visualizations for synthetic motions: The first row is the DTMs, the second row is the macro ChTMs, and the third row is the micro ChTMs.}
    \label{Synthetic_Visualizations}
    \vspace{-0.2cm}
\end{figure*}\par
\begin{figure*}[!ht]
    \centering
    \subfigure[]{
        \includegraphics[width=0.48\textwidth]{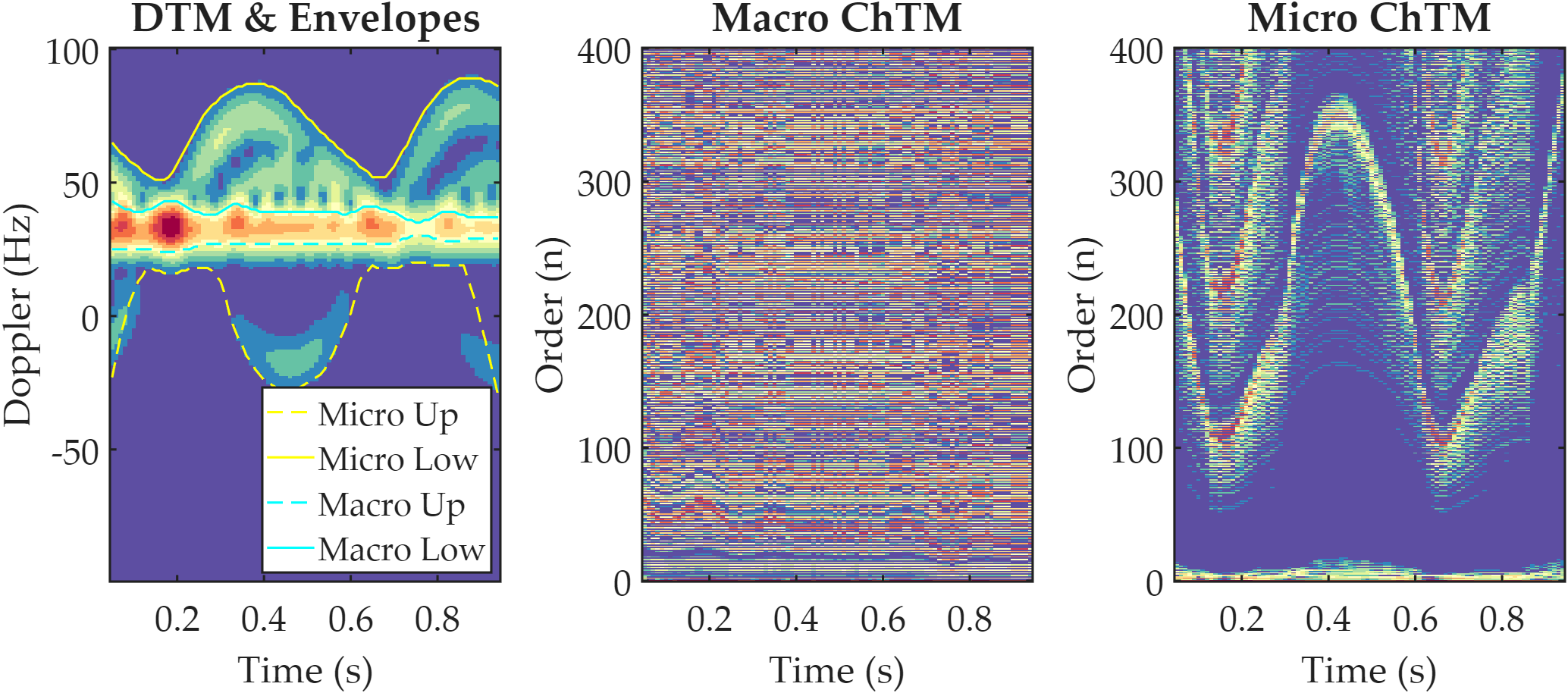}
        \label{fig:P1_Gun_Sim}
    }
    \hfil
    \subfigure[]{
        \includegraphics[width=0.48\textwidth]{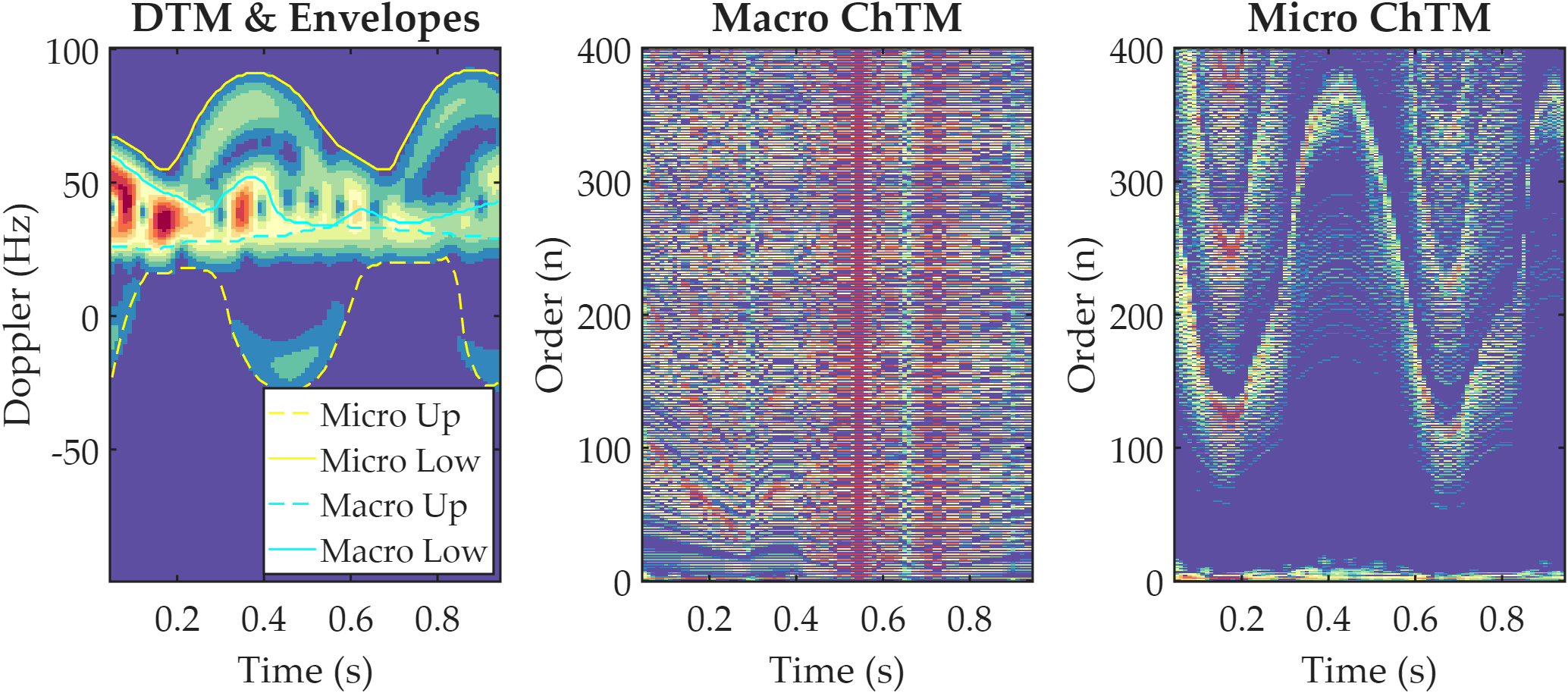}
        \label{fig:P1_Nogun_Sim}
    }
    \hfil
    \subfigure[]{
        \includegraphics[width=0.48\textwidth]{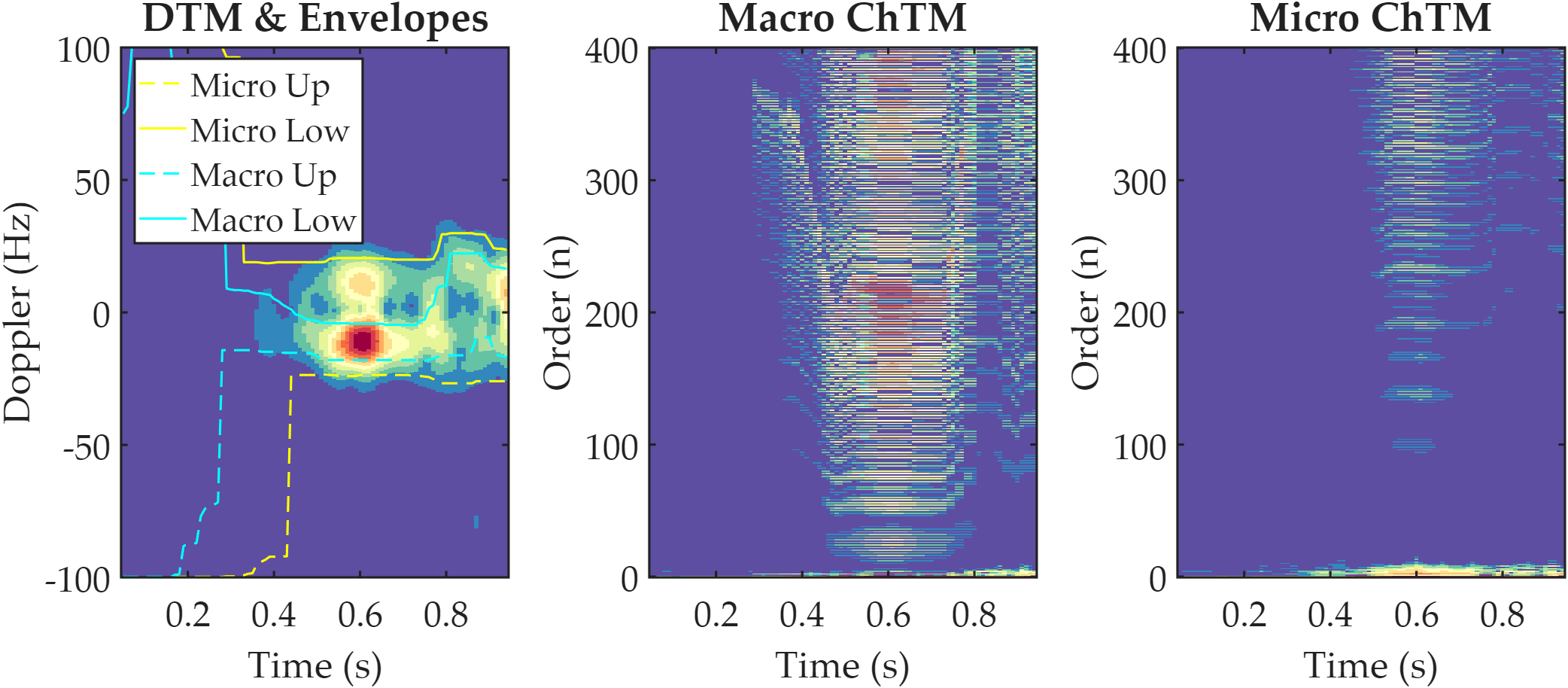}
        \label{fig:P1_Gun_Meas}
    }
    \hfil
    \subfigure[]{
        \includegraphics[width=0.48\textwidth]{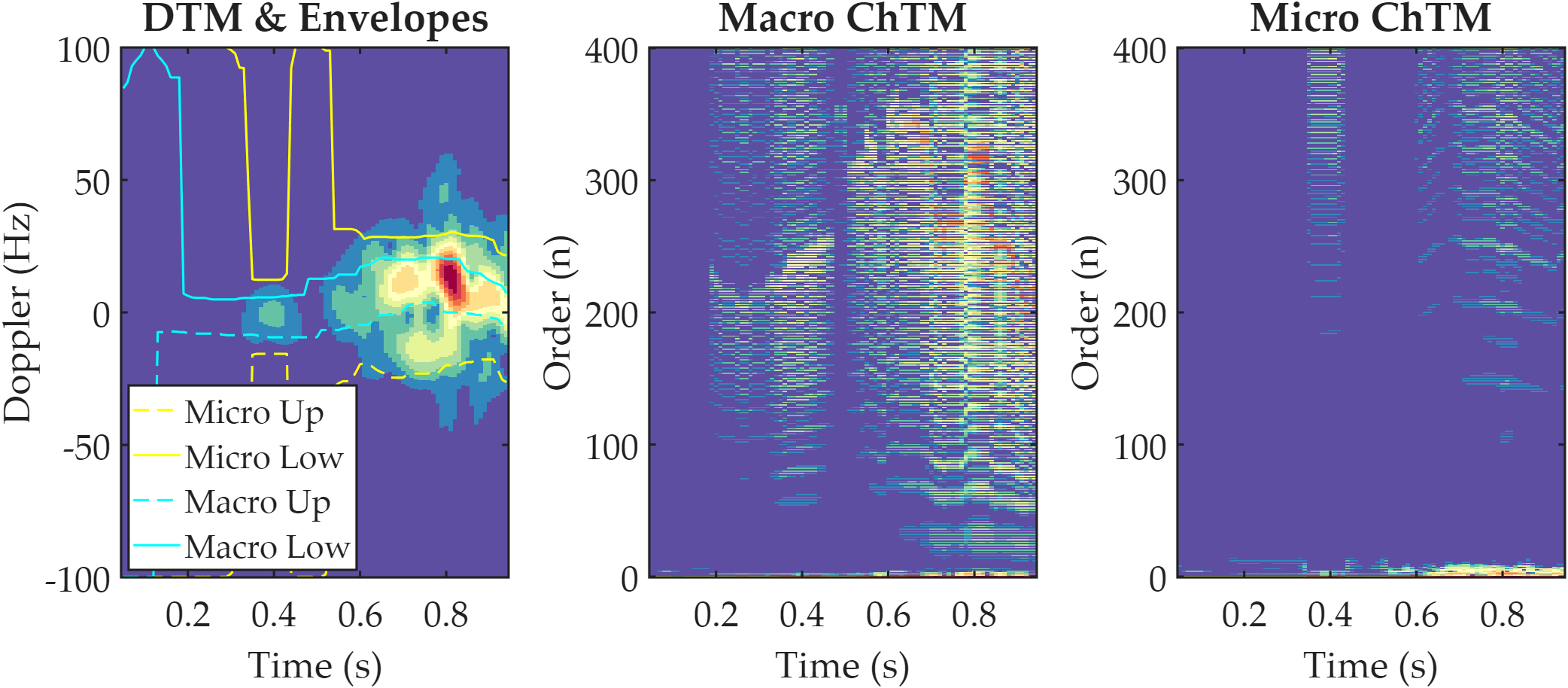}
        \label{fig:P1_Nogun_Meas}
    }
    \caption{Simulated and measured examples for tester P1: (a) Simulated result under armed condition, (b) Simulated result under unarmed condition, (c) Measured result under armed condition, and (d) Measured result under unarmed condition.}
    \label{Examples}
    \vspace{-0.3cm}
\end{figure*}\par
Neural network models are introduced in subsequent experiments, and uniform network training settings are presented in TABLE \ref{Training Settings}. Identical training hyperparameters are maintained for the proposed deep ensemble-based recognition method and the comparative network models. These parameters are characterized by a batch size of $32$, a total epoch number of $80$, and the utilization of the adaptive moment estimation optimizer with an initial learning rate of $0.0005$. Furthermore, the experiments are executed in an identical environment utilizing an NVIDIA Tesla V100 for training hardware, an NVIDIA RTX 3060 OC for validation hardware, and Paddlepaddle 3.0.0 in python 3.10 for software.\par

\subsection{Visualization Experiments}
As shown in Fig. \ref{Synthetic_Visualizations}, the visualizations of the generated data under specific motion patterns are presented to verify the characterization capability of the proposed method. The first row displays the traditional DTMs, the second row shows the macro ChTMs, and the third row presents the micro ChTMs, covering six distinct scenarios: acceleration, limp motion, complex harmonics, decelerating, fast running, and standard walking. From the figure, it can be seen that the micro-Doppler signature in the DTM accurately reflect the kinematic characteristics of different motions. The frequency fluctuation amplitude in fast running is observed to be significantly larger than that in standard walking, and the asymmetry in limp motion is clearly captured. Furthermore, the ChTMs in the third row exhibit strong structural correlation with the corresponding DTMs. The periodicity and energy distribution of the motions are effectively mapped into the order domain, where distinct textural patterns are formed for different motions. The results collectively demonstrate that the proposed ChTM effectively preserves the multi-order morphological detail information of the DTM and provides a robust feature representation for diverse limb motions.\par
\begin{figure*}
    \centering
    \includegraphics[width=\textwidth]{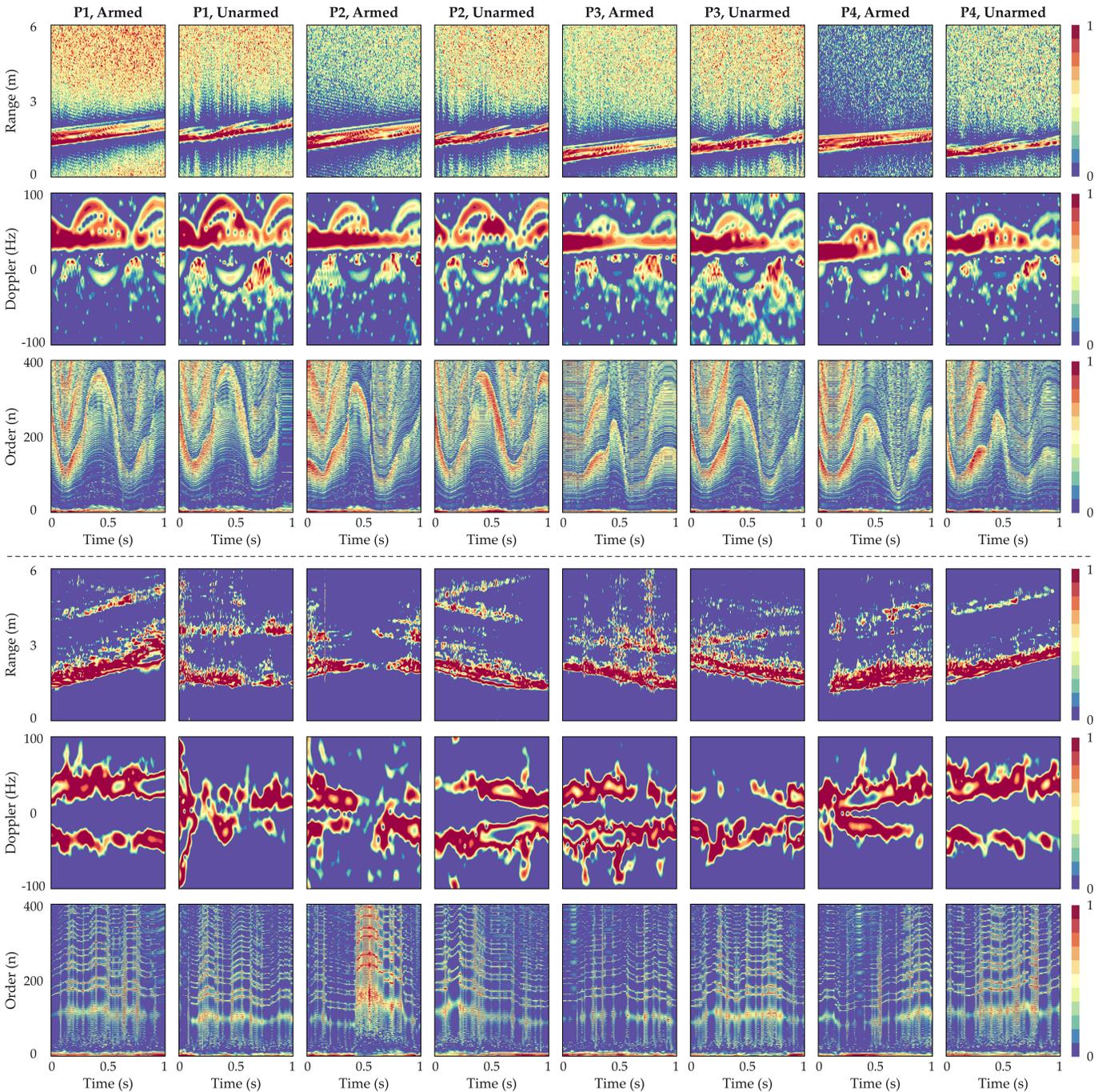}
    \caption{Visualizations for simulated and measured data: The first row is the simulated RTMs, the second row is the simulated DTMs, the third row is the simulated micro ChTMs, the forth row is the measured RTMs, the fifth row is the measured DTMs, and the sixth row is the measured micro ChTMs.}
    \label{Simulated_Measured_Visualizations}
    \vspace{-0.1cm}
\end{figure*}\par
As shown in Fig. \ref{Examples}, the envelope extraction results for tester P1 are presented to analyze the internal mechanism of the proposed method. The figure is divided into four sub-figures corresponding to simulated armed, simulated unarmed, measured armed, and measured unarmed conditions. In each sub-figure, the extracted envelopes overlaid on the DTM are displayed on the left, while the generated macro ChTM and micro ChTM are presented in the middle and on the right, respectively. From the DTMs, it is observed that the envelopes of the torso and limbs are accurately extracted. The distinctiveness between the micro-Doppler signature of armed and unarmed states is visualized through the variations in envelope amplitudes and trends. Furthermore, the time-frequency information is mapped into the Chebyshev coefficient space. The micro ChTM is observed to effectively aggregate the limb motion energy into specific polynomial orders, forming clear and distinguishable texture patterns. Although environmental noise and clutter are contained in the measured results, high structural consistency of the micro ChTM is maintained when compared to the simulated results. Consequently, the capability of the proposed method to extract and represent kinematic features from complex radar echoes is demonstrated.\par
As presented in Fig. \ref{Simulated_Measured_Visualizations}, a comprehensive comparison between the simulated data and the measured data is conducted to further verify the fidelity of the proposed method. The first three rows correspond to the simulated results, while the subsequent three rows correspond to the measured results. In each set, the RTMs, DTMs, and ChTMs are displayed in order. The columns represent different categories, denoted as P1 to P4 for the first tester to the forth tester, under armed and unarmed conditions. It is observed that the trajectory trends in the simulated RTMs are consistent with the measured data. The micro-Doppler signature in the simulated DTMs are shown to accurately replicate the limb motion details found in the real-world measurements. Furthermore, regarding the ChTMs in both third and sixth rows, distinct texture patterns and coefficient distributions are exhibited by different individuals and threat states, while consistency is maintained between the simulated and measured data for the same category. Consequently, the ChTM representation is verified to be capable of effectively characterizing human motion features and possessing certain class separability.\par
\begin{figure*}[!t]
    \centering
    \includegraphics[width=\textwidth]{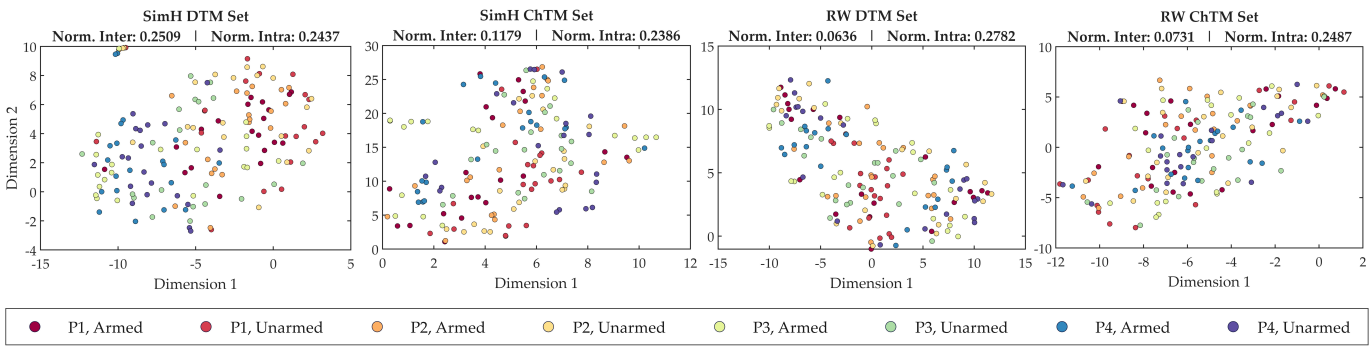}
    \caption{T-SNE results: The simulated DTM, the simulated ChTM, the measured DTM, and the measured ChTM are displayed from left to right, respectively.}
    \label{T-SNE}
    \vspace{-0.2cm}
\end{figure*}\par
\begin{table}[!t]
\begin{center}
\caption{Quantitative Comparison of T-SNE Performance$^{*}$.\label{TSNE_Metrics}}
\vspace{-0.2cm}
\resizebox{0.38\textwidth}{!}{
\begin{tabular}{ccccc}
\hline\hline
\multirow{2}{*}{\textbf{Metrics}} & \multicolumn{2}{c}{\textbf{SimH Dataset}} & \multicolumn{2}{c}{\textbf{RW Dataset}} \\
\cline{2-5} & \textbf{DTM} & \textbf{ChTM} & \textbf{DTM} & \textbf{ChTM} \\
\hline
Norm. Inter-Class$^{1}$ & $0.2509$ & $0.1179$ & $0.0636$ & $0.0731$ \\
Norm. Intra-Class$^{1}$ & $0.2437$ & $0.2386$ & $0.2782$ & $0.2487$ \\
\hline\hline
\end{tabular}
}
\par
\vspace{0.1cm}
\begin{minipage}{0.38\textwidth}
\footnotesize
$^{*}$ The results are also shown in the titles of Fig. \ref{T-SNE}.\\
$^{1}$ “Norm.” is the abbreviation of “Normalization”.
\end{minipage}
\end{center}
\vspace{-0.2cm}
\end{table}\par
As shown in both Fig. \ref{T-SNE} and TABLE \ref{TSNE_Metrics}, the feature distributions are visualized using the T-distributed stochastic neighbor embedding (T-SNE) algorithm \cite{T-SNE}, and the quantitative cluster separation metrics are calculated to evaluate the class separability of the proposed representation. To quantify the clustering performance, the normalized inter-class distance $D_{\mathrm{inter}}$ and the normalized intra-class distance $D_{\mathrm{intra}}$ are defined as:
\begin{equation}
\begin{aligned}
D_{\mathrm{inter}} &= \frac{2}{N_{\mathrm{cls}}(N_{\mathrm{cls}}-1)} \sum_{i=1}^{N_{\mathrm{cls}}-1} \sum_{j=i+1}^{N_{\mathrm{cls}}} \| \boldsymbol{\mu}_i - \boldsymbol{\mu}_j \|_2 \\
D_{\mathrm{intra}} &= \frac{1}{N_{\mathrm{cls}}} \sum_{i=1}^{N_{\mathrm{cls}}} \left( \frac{1}{N_{i}} \sum_{l=1}^{N_{i}} \| \mathbf{q}_{i,l} - \boldsymbol{\mu}_i \|_2 \right)
\end{aligned},
\end{equation}
where $N_{\mathrm{cls}}$ is the total number of activity categories, which is fixed as $8$ in our paper, $N_{i}$ is the number of samples in the $i$-th category, $\mathbf{q}_{i,l}$ is the coordinate vector of the $l$-th sample belonging to the $i$-th category in the reduced feature space, and $\boldsymbol{\mu}_i$ is the cluster center of the $i$-th category. The DTM feature spaces exhibit significant overlap and mixed distributions across different categories, corresponding to higher $D_{\mathrm{intra}}$ values. Conversely, distinct clusters with slightly clearer boundaries are formed in the ChTM feature spaces. It is observed from the quantitative results that the ChTM achieves lower $D_{\mathrm{intra}}$ and relatively stable $D_{\mathrm{inter}}$, which indicates that the proposed method yields certain class aggregation and superior separability compared to the traditional DTM representation.\par
\begin{figure*}[!t]
    \centering
    \includegraphics[width=\textwidth]{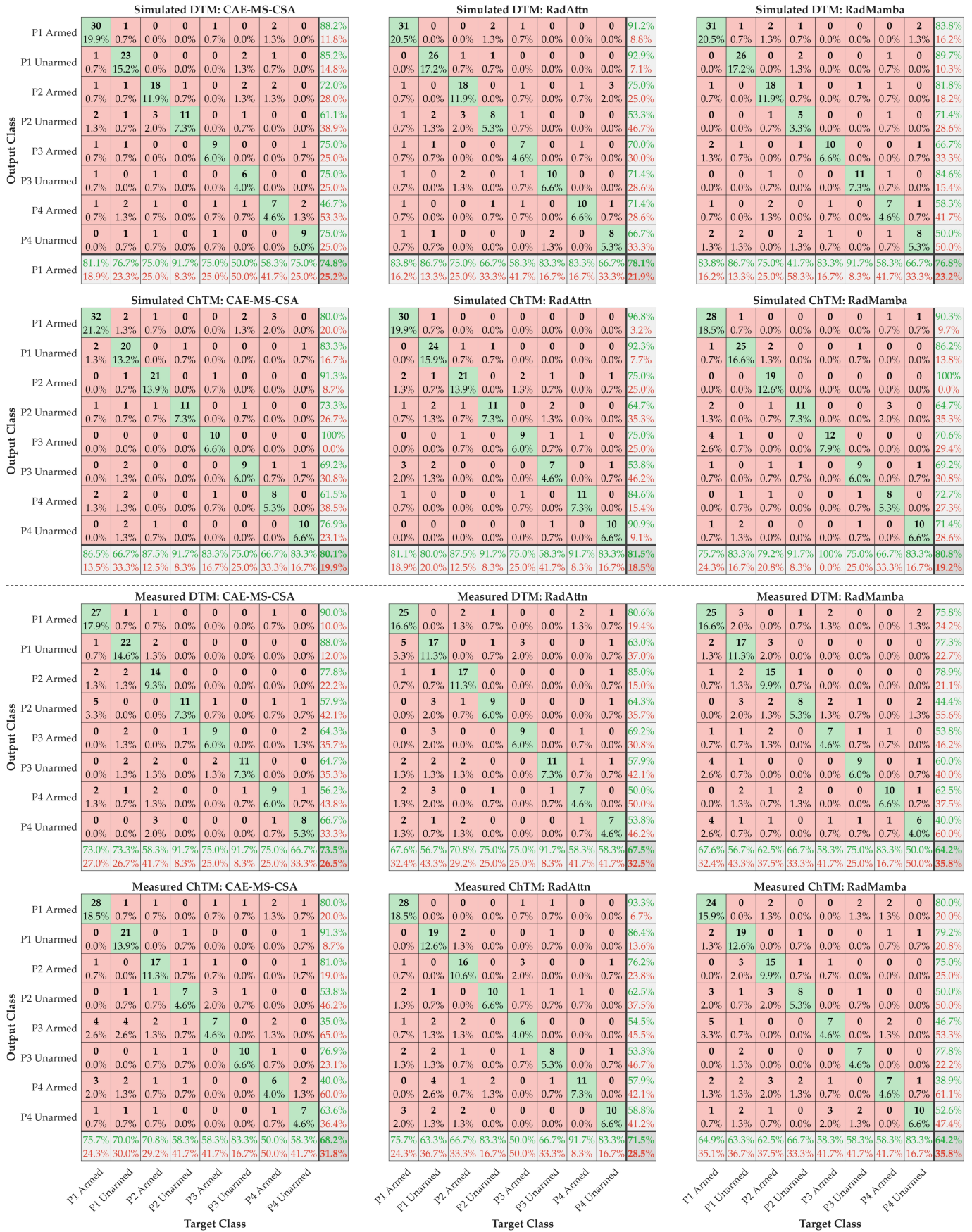}
    \caption{Confusion matrices: The simulated results are displayed in the first and the second rows, and the measured results are displayed in the third and the forth rows. All results are calculated based on the validation set.}
    \label{Confusion_Matrices}
    \vspace{-0.2cm}
\end{figure*}\par
\begin{table*}[!t]
\centering
\caption{Robustness Testing: Validation Accuracy Under Different Noise Levels$^{*}$.}
\label{Robustness_Testing}
\renewcommand{\arraystretch}{1.15}
\setlength{\tabcolsep}{3.5pt}
\scriptsize
\begin{tabular}{c|c|c|cccccccc|c}
\hline\hline
\multirow{2}{*}{\textbf{Method}} & \multirow{2}{*}{\textbf{Input}} & \multirow{2}{*}{\textbf{$\Delta$SNR (dB)}} & \multicolumn{8}{c|}{\textbf{Validation Accuracy per Class (\%)}} & \multirow{2}{*}{\textbf{OVA (\%)}} \\
\cline{4-11}
 & & & \textbf{P1-A} & \textbf{P1-U} & \textbf{P2-A} & \textbf{P2-U} & \textbf{P3-A} & \textbf{P3-U} & \textbf{P4-A} & \textbf{P4-U} & \\
\hline
\multicolumn{12}{c}{\textbf{Simulated Dataset}} \\
\hline
\multirow{10}{*}{CAE-MS-CSA} & \multirow{5}{*}{DTM} 
  & $0$ & $81.1$ & $76.7$ & $75.0$ & $91.7$ & $75.0$ & $50.0$ & $58.3$ & $75.0$ & $74.8$ \\
 & & $-4$ & $75.7$ & $70.0$ & $66.7$ & $83.3$ & $66.7$ & $41.7$ & $50.0$ & $66.7$ & $68.2$ \\
 & & $-8$ & $64.9$ & $60.0$ & $58.3$ & $75.0$ & $58.3$ & $33.3$ & $41.7$ & $58.3$ & $58.9$ \\
 & & $-12$ & $54.1$ & $50.0$ & $45.8$ & $58.3$ & $41.7$ & $25.0$ & $33.3$ & $41.7$ & $47.0$ \\
 & & $-16$ & $43.2$ & $40.0$ & $33.3$ & $41.7$ & $33.3$ & $16.7$ & $25.0$ & $25.0$ & $36.4$ \\
\cline{2-12}
 & \multirow{5}{*}{\textbf{ChTM}} 
  & $0$ & $86.5$ & $66.7$ & $87.5$ & $91.7$ & $83.3$ & $75.0$ & $66.7$ & $83.3$ & $\mathbf{80.1}$ \\
 & & $-4$ & $83.8$ & $63.3$ & $83.3$ & $83.3$ & $83.3$ & $66.7$ & $58.3$ & $83.3$ & $\mathbf{76.8}$ \\
 & & $-8$ & $78.4$ & $60.0$ & $79.2$ & $83.3$ & $75.0$ & $66.7$ & $58.3$ & $75.0$ & $\mathbf{72.8}$ \\
 & & $-12$ & $73.0$ & $56.7$ & $70.8$ & $75.0$ & $66.7$ & $58.3$ & $50.0$ & $66.7$ & $\mathbf{66.2}$ \\
 & & $-16$ & $67.6$ & $53.3$ & $62.5$ & $66.7$ & $58.3$ & $50.0$ & $41.7$ & $58.3$ & $\mathbf{59.6}$ \\
\hline
\multirow{10}{*}{RadarAttn} & \multirow{5}{*}{DTM} 
  & $0$ & $83.8$ & $86.7$ & $75.0$ & $66.7$ & $58.3$ & $83.3$ & $83.3$ & $66.7$ & $78.1$ \\
 & & $-4$ & $75.7$ & $76.7$ & $66.7$ & $58.3$ & $50.0$ & $75.0$ & $75.0$ & $58.3$ & $70.2$ \\
 & & $-8$ & $67.6$ & $66.7$ & $54.2$ & $50.0$ & $41.7$ & $66.7$ & $58.3$ & $50.0$ & $60.3$ \\
 & & $-12$ & $56.8$ & $53.3$ & $41.7$ & $41.7$ & $33.3$ & $50.0$ & $50.0$ & $33.3$ & $49.0$ \\
 & & $-16$ & $45.9$ & $40.0$ & $33.3$ & $25.0$ & $25.0$ & $33.3$ & $33.3$ & $25.0$ & $37.1$ \\
\cline{2-12}
 & \multirow{5}{*}{\textbf{ChTM}} 
  & $0$ & $81.1$ & $80.0$ & $87.5$ & $91.7$ & $75.0$ & $58.3$ & $91.7$ & $83.3$ & $\mathbf{81.5}$ \\
 & & $-4$ & $78.4$ & $76.7$ & $83.3$ & $83.3$ & $75.0$ & $58.3$ & $83.3$ & $75.0$ & $\mathbf{78.1}$ \\
 & & $-8$ & $73.0$ & $73.3$ & $79.2$ & $83.3$ & $66.7$ & $50.0$ & $75.0$ & $75.0$ & $\mathbf{73.5}$ \\
 & & $-12$ & $67.6$ & $66.7$ & $70.8$ & $75.0$ & $58.3$ & $50.0$ & $66.7$ & $66.7$ & $\mathbf{67.5}$ \\
 & & $-16$ & $62.2$ & $60.0$ & $62.5$ & $66.7$ & $50.0$ & $41.7$ & $58.3$ & $58.3$ & $\mathbf{60.3}$ \\
\hline
\multirow{10}{*}{RadMamba} & \multirow{5}{*}{DTM} 
  & $0$ & $83.8$ & $86.7$ & $75.0$ & $41.7$ & $83.3$ & $91.7$ & $58.3$ & $66.7$ & $76.8$ \\
 & & $-4$ & $75.7$ & $76.7$ & $66.7$ & $33.3$ & $75.0$ & $83.3$ & $50.0$ & $58.3$ & $68.9$ \\
 & & $-8$ & $64.9$ & $66.7$ & $58.3$ & $25.0$ & $58.3$ & $66.7$ & $41.7$ & $41.7$ & $58.3$ \\
 & & $-12$ & $54.1$ & $56.7$ & $45.8$ & $16.7$ & $50.0$ & $50.0$ & $33.3$ & $33.3$ & $48.3$ \\
 & & $-16$ & $43.2$ & $43.3$ & $33.3$ & $16.7$ & $33.3$ & $33.3$ & $25.0$ & $16.7$ & $37.7$ \\
\cline{2-12}
 & \multirow{5}{*}{\textbf{ChTM}} 
  & $0$ & $75.7$ & $83.3$ & $79.2$ & $91.7$ & $100.0$ & $75.0$ & $66.7$ & $83.3$ & $\mathbf{80.8}$ \\
 & & $-4$ & $73.0$ & $80.0$ & $75.0$ & $83.3$ & $91.7$ & $66.7$ & $66.7$ & $75.0$ & $\mathbf{77.5}$ \\
 & & $-8$ & $67.6$ & $76.7$ & $70.8$ & $75.0$ & $83.3$ & $66.7$ & $58.3$ & $66.7$ & $\mathbf{72.2}$ \\
 & & $-12$ & $62.2$ & $70.0$ & $66.7$ & $66.7$ & $75.0$ & $58.3$ & $50.0$ & $58.3$ & $\mathbf{66.2}$ \\
 & & $-16$ & $56.8$ & $63.3$ & $58.3$ & $58.3$ & $66.7$ & $50.0$ & $50.0$ & $50.0$ & $\mathbf{59.6}$ \\
\hline
\multicolumn{12}{c}{\textbf{Measured Dataset}} \\
\hline
\multirow{10}{*}{CAE-MS-CSA} & \multirow{5}{*}{DTM} 
  & $0$ & $73.0$ & $73.3$ & $58.3$ & $91.7$ & $75.0$ & $91.7$ & $75.0$ & $66.7$ & $73.5$ \\
 & & $-4$ & $64.9$ & $63.3$ & $50.0$ & $83.3$ & $66.7$ & $83.3$ & $66.7$ & $58.3$ & $64.9$ \\
 & & $-8$ & $54.1$ & $53.3$ & $41.7$ & $66.7$ & $50.0$ & $66.7$ & $50.0$ & $41.7$ & $53.6$ \\
 & & $-12$ & $43.2$ & $40.0$ & $33.3$ & $50.0$ & $41.7$ & $50.0$ & $41.7$ & $25.0$ & $41.7$ \\
 & & $-16$ & $32.4$ & $30.0$ & $25.0$ & $33.3$ & $25.0$ & $33.3$ & $25.0$ & $16.7$ & $29.8$ \\
\cline{2-12}
 & \multirow{5}{*}{\textbf{ChTM}} 
  & $0$ & $75.7$ & $70.0$ & $70.8$ & $58.3$ & $58.3$ & $83.3$ & $50.0$ & $58.3$ & $\mathbf{68.2}$ \\
 & & $-4$ & $73.0$ & $66.7$ & $66.7$ & $58.3$ & $58.3$ & $75.0$ & $50.0$ & $50.0$ & $\mathbf{65.6}$ \\
 & & $-8$ & $67.6$ & $63.3$ & $62.5$ & $50.0$ & $50.0$ & $66.7$ & $41.7$ & $41.7$ & $\mathbf{60.3}$ \\
 & & $-12$ & $62.2$ & $56.7$ & $58.3$ & $41.7$ & $41.7$ & $58.3$ & $41.7$ & $41.7$ & $\mathbf{55.6}$ \\
 & & $-16$ & $56.8$ & $50.0$ & $50.0$ & $41.7$ & $33.3$ & $50.0$ & $33.3$ & $33.3$ & $\mathbf{49.7}$ \\
\hline
\multirow{10}{*}{RadarAttn} & \multirow{5}{*}{DTM} 
  & $0$ & $67.6$ & $56.7$ & $70.8$ & $75.0$ & $75.0$ & $91.7$ & $58.3$ & $58.3$ & $67.5$ \\
 & & $-4$ & $59.5$ & $50.0$ & $62.5$ & $66.7$ & $66.7$ & $83.3$ & $50.0$ & $50.0$ & $59.6$ \\
 & & $-8$ & $48.6$ & $40.0$ & $50.0$ & $50.0$ & $50.0$ & $66.7$ & $41.7$ & $41.7$ & $48.3$ \\
 & & $-12$ & $37.8$ & $30.0$ & $41.7$ & $41.7$ & $33.3$ & $50.0$ & $33.3$ & $25.0$ & $37.1$ \\
 & & $-16$ & $27.0$ & $23.3$ & $29.2$ & $25.0$ & $25.0$ & $33.3$ & $25.0$ & $16.7$ & $26.5$ \\
\cline{2-12}
 & \multirow{5}{*}{\textbf{ChTM}} 
  & $0$ & $75.7$ & $63.3$ & $66.7$ & $83.3$ & $50.0$ & $66.7$ & $91.7$ & $83.3$ & $\mathbf{71.5}$ \\
 & & $-4$ & $73.0$ & $60.0$ & $62.5$ & $75.0$ & $50.0$ & $66.7$ & $83.3$ & $75.0$ & $\mathbf{68.2}$ \\
 & & $-8$ & $67.6$ & $56.7$ & $58.3$ & $66.7$ & $41.7$ & $58.3$ & $75.0$ & $66.7$ & $\mathbf{62.9}$ \\
 & & $-12$ & $62.2$ & $53.3$ & $54.2$ & $58.3$ & $41.7$ & $50.0$ & $66.7$ & $58.3$ & $\mathbf{57.6}$ \\
 & & $-16$ & $56.8$ & $46.7$ & $50.0$ & $50.0$ & $33.3$ & $41.7$ & $50.0$ & $50.0$ & $\mathbf{51.7}$ \\
\hline
\multirow{10}{*}{RadMamba} & \multirow{5}{*}{DTM} 
  & $0$ & $67.6$ & $56.7$ & $62.5$ & $66.7$ & $58.3$ & $75.0$ & $83.3$ & $50.0$ & $64.2$ \\
 & & $-4$ & $59.5$ & $50.0$ & $54.2$ & $58.3$ & $50.0$ & $66.7$ & $66.7$ & $41.7$ & $56.3$ \\
 & & $-8$ & $48.6$ & $40.0$ & $41.7$ & $41.7$ & $41.7$ & $50.0$ & $50.0$ & $33.3$ & $45.0$ \\
 & & $-12$ & $37.8$ & $30.0$ & $33.3$ & $33.3$ & $25.0$ & $33.3$ & $33.3$ & $25.0$ & $33.8$ \\
 & & $-16$ & $29.7$ & $20.0$ & $25.0$ & $16.7$ & $16.7$ & $25.0$ & $25.0$ & $16.7$ & $24.5$ \\
\cline{2-12}
 & \multirow{5}{*}{\textbf{ChTM}} 
  & $0$ & $64.9$ & $63.3$ & $62.5$ & $66.7$ & $58.3$ & $58.3$ & $58.3$ & $83.3$ & $\mathbf{64.2}$ \\
 & & $-4$ & $62.2$ & $60.0$ & $58.3$ & $58.3$ & $50.0$ & $50.0$ & $50.0$ & $75.0$ & $\mathbf{60.3}$ \\
 & & $-8$ & $56.8$ & $53.3$ & $54.2$ & $50.0$ & $41.7$ & $41.7$ & $41.7$ & $66.7$ & $\mathbf{54.3}$ \\
 & & $-12$ & $48.6$ & $46.7$ & $45.8$ & $41.7$ & $41.7$ & $41.7$ & $33.3$ & $58.3$ & $\mathbf{47.0}$ \\
 & & $-16$ & $43.2$ & $40.0$ & $41.7$ & $33.3$ & $33.3$ & $33.3$ & $25.0$ & $41.7$ & $\mathbf{40.4}$ \\
\hline\hline
\multicolumn{12}{l}{\footnotesize $^{*}$ Abbreviations: P1-A (P1 Armed), P1-U (P1 Unarmed), P2-A (P2 Armed), P2-U (P2 Unarmed), etc. OVA (Overall Validation Accuracy).}
\end{tabular}
\vspace{-0.2cm}
\end{table*}\par

\subsection{Comparison Experiments}
As shown in Fig. \ref{Confusion_Matrices}, the validation confusion matrices are displayed to evaluate the recognition performance of both DTM based and ChTM based micro-Doppler representation. The figure is arranged into four rows and three columns, where the first two rows correspond to the simulated dataset and the last two rows correspond to the measured dataset. Furthermore, three different existing network models developed for radar based HAR published in early 2026, including CAE-MS-CSA \cite{CAE-MS-CSA}, RadarAttn \cite{RadarAttn}, and RadMamba \cite{RadMamba}, are employed for comparison in the column direction. It is observed from the DTM based results that significant confusion exists between the armed and unarmed states for the same identity, which indicates that the subtle kinematic differences are difficult to distinguish using traditional time-frequency features. In contrast, when the proposed ChTM is utilized as the input, the diagonal dominance of the confusion matrices is slightly enhanced, and the misclassification errors between similar categories are suppressed in certain degree. The overall validation accuracy between DTM based and ChTM based micro-Doppler representation is close. The results are consistently exhibited across both simulated and measured datasets and among different network architectures. Consequently, the effectiveness of the proposed ChTM in characterizing fine-grained micro-Doppler signature is demonstrated.\par
As listed in TABLE \ref{Robustness_Testing}, the robustness of the proposed method is evaluated by introducing additive white Gaussian noise with different signal-to-noise ratio (SNR) levels to the raw radar echoes \cite{Robustness}. The overall validation accuracy and the class-wise accuracy for both simulated and measured datasets are recorded to assess the performance degradation trends. It is observed that the recognition accuracy of all methods exhibits a declining trend as the SNR decreases from $0\mathrm{~dB}$ to $-16\mathrm{~dB}$. However, compared with the DTM based representation, the ChTM based methods demonstrate stronger noise immunity. For instance, in the measured dataset using the RadarAttn backbone, when the $\Delta$SNR drops to $-16\mathrm{~dB}$, the overall accuracy of the DTM decreases rapidly to $26.5\%$, whereas the ChTM maintains a relatively high accuracy of $51.7\%$. This phenomenon indicates that the envelope extraction and orthogonal projection mechanism of the ChTM generation effectively filters out some random noise while preserving the topological structure of human motion. Consequently, the robustness of the proposed representation is verified.\par
\begin{table*}[!t]
\centering
\caption{Results of Different Chebyshev Order $N_{\mathrm{order}}$ on Validation Accuracy.}
\vspace{-0.2cm}
\label{Sensitivity_Analysis}
\renewcommand{\arraystretch}{1.15}
\setlength{\tabcolsep}{8pt}
\scriptsize
\begin{tabular}{c|c|cccccc}
\hline\hline
\multirow{2}{*}{\textbf{Dataset}} & \multirow{2}{*}{\textbf{Method}} & \multicolumn{6}{c}{\textbf{Validation Accuracy with Different Orders (\%)}} \\
\cline{3-8} 
 &  & $N\!=\!4$ & $N\!=\!8$ & $N\!=\!16$ & $N\!=\!32$ & $N\!=\!48$ & $N\!=\!64$ \\
\hline
\multirow{3}{*}{Simulated} 
 & CAE-MS-CSA & $38.4$ & $65.6$ & $75.5$ & $80.1$ & $\mathbf{80.8}$ & $78.8$ \\
 & RadarAttn  & $47.7$ & $63.6$ & $77.5$ & $81.5$ & $\mathbf{82.1}$ & $81.5$ \\
 & RadMamba   & $47.0$ & $66.2$ & $76.2$ & $\mathbf{80.8}$ & $80.1$ & $79.5$ \\
\hline
\multirow{3}{*}{Measured} 
 & CAE-MS-CSA & $39.7$ & $54.3$ & $64.2$ & $\mathbf{68.2}$ & $\mathbf{68.2}$ & $67.5$ \\
 & RadarAttn  & $43.0$ & $56.3$ & $66.9$ & $\mathbf{71.5}$ & $\mathbf{71.5}$ & $70.9$ \\
 & RadMamba   & $39.1$ & $50.3$ & $60.3$ & $\mathbf{64.2}$ & $\mathbf{64.2}$ & $62.9$ \\
\hline\hline
\multicolumn{8}{l}{\footnotesize $^{*}$ The baseline results at $0\mathrm{~dB}$ $\Delta$SNR reported in previous experiments correspond to $N_{\mathrm{order}}=32$.}
\end{tabular}
\vspace{-0.2cm}
\end{table*}\par
As presented in TABLE \ref{Sensitivity_Analysis}, the sensitivity analysis of the Chebyshev order is conducted to investigate the impact of feature dimensions on recognition performance. It is observed that when the order is set to a low value, such as $4$ or $8$, the validation accuracy is relatively poor for both simulated and measured datasets. As the order increases from $8$ to $48$, improvement in recognition accuracy is exhibited across all network models. The performance tends to saturate or reach a peak when the order reaches $48$ or $64$. However, when the order is further increased to $64$, a slight decline or stagnation in accuracy is observed in certain cases, particularly for the measured dataset. This phenomenon indicates that excessively high orders may introduce high-frequency noise components or lead to feature redundancy, which negatively affects the recognition capability of the classifier. The optimal order selection is inconsistent with the conclusion derived from the theory. This discrepancy arises because the human body structure in actual detection scenarios is more complex than the model, and is also affected by noise. However, the overall trend conforms to physical laws. Consequently, considering the trade-off between recognition accuracy and data compression efficiency, the order of $32$ is selected as the optimal setting in this paper, as it provides robust feature representation with reduced dimension compared to the original DTM.\par

\subsection{Discussions}
Based on the numerical simulations and experimental results presented above, the performance characteristics of the proposed method are discussed from three aspects: The mechanism of robustness, the selection of the optimal polynomial order, and the practical application value.\par
First, the robustness of the ChTM stems from the orthogonal projection, which functions as a feature filter. Since human kinematic features are concentrated in low-to-mid orders while random noise is distributed across high orders, truncating the decomposition order effectively suppresses noise while preserving the topological structure of micro-Doppler signature.\par
Second, the experimentally optimal order ($N_{\mathrm{order}}=32$) exceeds the theoretical lower bound ($N_{\mathrm{req}} \approx 10$) derived from the ideal point-scatterer model. This discrepancy arises because real-world human targets exhibit complex extended scattering and random complex micro-motions, necessitating higher orders for accurate reconstruction. However, excessively high orders introduce high-frequency clutter. Thus, $N_{\mathrm{order}}=32$ represents a physical trade-off between feature fidelity and noise suppression.\par
Finally, the proposed method achieves data compression, but fails to maintain excellent accuracy. However, by reducing the input dimension from hundreds of frequency bins to only $32$ coefficients per time step, the computational burden and memory footprint are minimized. If a model can be developed to recognize the ChTM without resizing it into a square, its real-time performance can be effectively improved.\par

%%%%%%%%%%%%%%%%%%%%%%%%%%%%%%%%%%%%%%%%%%%%%%%%%%%%%%%%%%%%%%%%%%%%%%%%%%%%%%%%%%
% Conclusion
%%%%%%%%%%%%%%%%%%%%%%%%%%%%%%%%%%%%%%%%%%%%%%%%%%%%%%%%%%%%%%%%%%%%%%%%%%%%%%%%%%
\section{Conclusion}
In this paper, the ChTM has been proposed to address the challenges of minimal micro-Doppler distinctiveness and low inference efficiency in TWR HAR. Parametric kinematic models and the radar echo model have been established. A feature representation method based on orthogonal Chebyshev polynomial decomposition has been developed, where kinematic envelopes have been extracted and time-frequency slices have been mapped into a robust coefficient space. Through numerical simulations and experiments, the effectiveness of the proposed method has been verified. It has been demonstrated that armed and unarmed activities have been effectively characterized by the proposed method, and a balance between recognition accuracy and input data dimensions has been achieved through spectrum compression.\par

%%%%%%%%%%%%%%%%%%%%%%%%%%%%%%%%%%%%%%%%%%%%%%%%%%%%%%%%%%%%%%%%%%%%%%%%%%%%%%%%%%
% Acknowledgment
%%%%%%%%%%%%%%%%%%%%%%%%%%%%%%%%%%%%%%%%%%%%%%%%%%%%%%%%%%%%%%%%%%%%%%%%%%%%%%%%%%
\section{Acknowledgment}
I would like to thank my parents, my sweetheart, and my mentor professor Xiaopeng Yang. It is their encouragement and support that have motivated me to improve this research. In addition, I would also like to thank Dr. Jiarong Zhao, my fellow colleague. I had deep discussions with him about the theoretical model, and he gave me some inspiration.\par
No idea is useless, even if it is just as simple as a transformation, and even if the results are not all state-of-the-art. Guided by this belief, this work came into being. In addition, it is hoped that this new map representation form can really be put into use.\par

%%%%%%%%%%%%%%%%%%%%%%%%%%%%%%%%%%%%%%%%%%%%%%%%%%%%%%%%%%%%%%%%%%%%%%%%%%%%%%%%%%
% References
%%%%%%%%%%%%%%%%%%%%%%%%%%%%%%%%%%%%%%%%%%%%%%%%%%%%%%%%%%%%%%%%%%%%%%%%%%%%%%%%%%

\end{document}